\documentstyle[12pt,epsf,aaspp4]{article}

\def\msun{{\rm\,M_\odot}}

\newcommand{\etal}{et al.\ }

\newcommand{\lya}{Ly$\alpha$ }

\newcommand{\sci}[1]{{\rm \; \times \; 10^{#1}}}
\newcommand{\cm}[1]{\, {\rm cm^{#1}}}
\newcommand{\N}[1]{{N({\rm #1})}}

\def\etal   {{et~al.}\ }
\def\zsun{{\rm\,Z_\odot}}
\def\msun{{\rm\,M_\odot}}

\def\vol#1  {{{#1}{\rm,}\ }}

\def\etal{et al.\ }

\begin{document}
\title{Metallicity Evolution of Damped Lyman Alpha Systems In $\Lambda$CDM Cosmology}
\author{Renyue Cen\altaffilmark{1}, Jeremiah P. Ostriker\altaffilmark{2},
Jason X. Prochaska\altaffilmark{3} and Arthur M. Wolfe\altaffilmark{4}}
\altaffiltext{1} {Princeton University Observatory, Princeton University, Princeton, NJ 08544; cen@astro.princeton.edu}
\altaffiltext{2} {Princeton University Observatory, Princeton University, Princeton, NJ 08544; jpo@astro.princeton.edu}
\altaffiltext{3} {The Observatories of the Carnegie Institute of Washington, 813 Santa Barbara St., Pasadena, CA 91101; xavier@ociw.edu}
\altaffiltext{4} {Department of Physics and Center for Astrophysical and Space Sciences, University of California, San Diego, C-0424, La Jolla, CA 92093; art@kingpin.ucsd.edu}

\begin{abstract}
Utilizing a new, high mass resolution ($\Delta m_b=10^{5.5}\msun$)
hydrodynamic simulation of a spatially flat $\Lambda$CDM cosmological model
with detailed microphysics and galaxy formation,
including radiation shielding,
energy deposition
and metal enrichment 
from supernovae and associated metal cooling/heating,
we compute the metallicity evolution of damped Lyman alpha systems (DLAs) and 
find a reasonable agreement with observations.
In particular, the observed slow evolution of 
the DLA metallicity occurs naturally in the simulation
due to the combined effects of physical and observational
selection.
The slow metallicity evolution is caused 
by the steady transformation, with increasing time,
of the highest metallicity systems to ``galaxies",
thus depleting this category, while all the lower
metallicity systems show, individually,
an increase in metallicity.
Although the trend of DLA metallicity %(column density weighted)
with redshift is in good agreement with observations,
it appears that the average metallicity of simulated
DLAs is higher than observed by $0.3-0.5$ dex in the probed
redshift range ($z=0-5$).
Our study indicates that this difference 
may be attributed to observational selection effects due to
dust obscuration.
If we allow for a dust obscuration effect,
our model reproduces the observed metallicity 
evolution in both amplitude
and slope.

We find that DLAs 
are not a simple population but probe
a range of different systems 
and the mix changes with redshift.
The median luminosity of a DLA, $L_{DLA}(z)$,
in units of typical galaxy luminosity  at that redshift,
$L^*(z)$, that is, $(L_{DLA}/L^*(z))_z$, decreases
from $1.1$ to $0.5$ as redshift declines from $z=3$ to $z=0$,
but the absolute luminosity of the median DLA system
increases in the same interval by a
factor of five from 
$0.1L^*(z=0)$
to 
$0.5L^*(z=0)$.
About 50\% of all metals in the gaseous phase
is in DLAs at all times from $z=5$ to $z=1$,
making a rapid downturn at $z\le 1$ to $\sim 20\%$ by $z=0$,
as metals are swept into the hotter components of the IGM
as well as locked up in stars.

While not the primary focus of this study,
we find that the model provides good matches to
observations with respect to column density distribution
and evolution of neutral gas content, if the same dust obscuration is 
taken into account.
We find $\Omega_{DLA,comp}=1-3\times 10^{-3}$, depending on
the effect of dust obscuration.
%However, the model may predict too many damped
%Lyman alpha systems at low column densities,
%indicating either imperfections of the simulation,
%or, more interestingly,
%the possibility that this 
%$\Lambda$CDM model
%has too much small scale power.

\end{abstract}

\keywords{Cosmology: large-scale structure of Universe 
-- cosmology: theory
-- intergalactic medium 
-- quasars: absorption lines 
-- hydrodynamics}

\section{Introduction}

Since their discovery, it has been recognized that studies of the
damped \lya systems help reveal the chemical enrichment history
of the universe in neutral gas.  
These quasar absorption line systems,
defined to exhibit $\N{HI}\ge2\sci{20} \cm{-2}$, dominate the neutral
hydrogen content of the universe 
up to $z\ge 4$ (Storrie-Lombardi and Wolfe 2000;
Peroux \etal 2002)
and exhibit metal-line profiles which yield accurate metallicity
measurements (e.g., Prochaska et al. 2001).
The first comprehensive metallicity survey of the damped systems 
(Pettini \etal 1994)
revealed a sub-solar metallicity 
in systems at $z \sim 2$ with a considerable scatter from system to
system.  
By weighting these individual Zn metallicity measurements
by the $\N{HI}$ value of the system, the authors derived a global 
mean metallicity $\log(Z_{HI}/Z_\odot) \approx -1.1$~dex.  
This important quantity is independent of the morphology, age, 
and history of the individual damped \lya systems.
Further observations 
(Pettini \etal 1997)
supported the initial work and 
suggested minimal evolution in $Z_{HI}$ from $z \sim 1 - 2.5$ 
(Pettini 1999).

Prochaska \& Wolfe (2000a)
extended metallicity surveys of the damped systems to
$z \sim 4.5$ through echelle observations on the Keck~I 10m telescope.
Their results also favored minimal evolution in the metallicity of HI gas
from $z \sim 2 - 4$, although their initial sample suffered from 
small number statistics.  More recently, 
Prochaska, Gawiser, \& Wolfe (2001)
and 
Prochaska \& Wolfe (2002)
have enlarged the sample to $N \approx 50$ systems at $z\ge 2$ and have
confirmed the main conclusions from Prochaska \& Wolfe (2000a):
(1) the $\N{HI}$-weighted
mean metallicity exhibits minimal evolution from $z \sim 1.7 - 3.5$ and
a small but significant decline at $z \ge 3.5$;
(2) no individual damped system shows a metallicity below 1/1000 solar;
and 
(3) the unweighted mean metallicity closely tracks the weighted mean over
this entire redshift range.
%[Mention Songaila and Cowie result at $z \sim 5$?]

At low redshift $(z \le 1$), the situation is far less certain.  There
are many fewer observations (Boiss\'e \etal 1998; Pettini \etal 2000)
and the majority of
systems have very low $\N{HI}$ values such that the weighted mean is
entirely dominated by one or two DLAs.  Although the current
value for $Z_{HI}$ at $z \le 1$ roughly matches the $z \ge 2$ values, it is 
very possible that this is the result of small number statistics
(Kulkarni \& Fall 2002).

On the theoretical front, 
various groups have composed models to predict the chemical evolution
history of the universe.  The majority of these 
(Pei \& Fall 1995;
Pei \etal 1999;
Malaney \& Chaboyer 1996;
Edmunds \& Phillips 1997)
compute the evolution of gas metallicity globally.
All such calculations have been constrained to yield $Z_{HI} = Z_\odot$
at $z=0$ and all predict marked evolution in $Z_{HI}$ from $z=4$ to 
the present epoch.  In comparison with the observations, 
these scenarios predict more substantial evolution in $Z_{HI}$ at high redshift
than is currently observed.  
As our subsequent results would indicate,
this global approach does not accurately reflect 
the unique evolution of the DLA sub population.
Somerville, Primack, \& Faber (2001) used semi-analytic models
to derive metallicity evolution and find reasonable agreement
with observations.
It is, however, difficult to gauge whether the lack of treatment
of the intergalactic medium in the semi-analytic models
has substantially affected the metallicity in DLAs.

We have followed a direct approach using cosmological hydrodynamic
simulations with galaxy formation. 
In an earlier paper (Cen \& Ostriker 1999)
we attempted to compute cosmic chemical evolution
using a simulation 
with a baryonic mass resolution of $1\times 10^8h^{-1}\msun$
and a spatial resolution of $\sim 200h^{-1}$kpc comoving.
While the agreement with observations found in a variety
of environments ranging 
from rich clusters of galaxies
to low column density Lyman alpha clouds
was encouraging, we found ourselves unable to adequately address
the evolution of metallicity of dense and small
systems such as DLAs.
Tissera \etal (2001) also used direct (SPH) simulations to 
study the chemical evolution of DLAs but their 
simulations suffered substantially from limited mass resolution.
Since DLAa are the prime repository of cool gas phase
baryons, we thought it important to study them using
a new simulation especially designed for this purpose.
Our new simulation has 
a baryonic mass resolution of $3.5\times 10^5h^{-1}\msun$ and
a spatial resolution of $\sim 30h^{-1}$kpc comoving,
and we now investigate 
%in detail
the evolution of metallicity of DLAs.
The central questions that we would like to address are:
Why is the observed evolution of metallicity 
of DLAs so mild from $z=4-0$, while we know that 
the star formation primarily occurs during this redshift interval?
What is 
the expected distribution of metallicity among DLAs?
How does the evolution of metallicity of DLAs
compare to that of the all gas in the universe globally averaged
and to stellar metallicity?
In what sense (if any) are DLAs progenitors of normal galaxies?
A description of the simulation is given in \S 2.
Results are presented defined in \S 3,
followed by conclusions in \S 4.

We would like to point out at outset 
that we are interested in ``cosmic chemical evolution" of DLAs,
which is a convolution of true galactic
evolution and observational selection of DLAs.
This has a different
meaning for investigators who work in galactic studies where
the same term means internal galactic chemical evolution 
with some chosen boundary conditions.
Our simulations are admittedly inadequate 
for explaining any detailed chemical
evolution of individual galaxies,
for which studies such as that  
by Mathlin \etal (2001) can make much finer
predictions to compare with respective observations.
But, as they pointed out, galaxy interaction and merger,
and gas infall/outflow (which 
are known to be important processes
in hierarchical structure formation)
may have played a major role but were ignored in their treatment.
Our simulations, on the other hand, take into account
mergers, gas infall/outflow and other galaxy interactions
and thus provide a complementary approach to this problem.

\section{Simulations}

The results reported on here are based on a new computation of 
the evolution of the intergalactic medium in 
a cold dark matter model with a cosmological constant.
The relevant model parameters
are: $\Omega_0=0.30$, $\Omega_b=0.035$, $\Lambda_0=0.70$, $\sigma_8=0.90$,
$H_0=67$km/s/Mpc and $n=1.0$.
The cosmological model
is normalized to both the cosmic microwave background (CMB) temperature
fluctuations measured by COBE on large scales (e.g., Bunn \& White 1997)
and the observed abundance of clusters of galaxies in
the local universe (e.g., Cen 1998), and
it is close to both the concordance model of Ostriker \& Steinhardt (1995) 
and consistent with the recent
high redshift supernova results (Reiss \etal 1998)
and CMB measurement on intermediate scales
(de Bernardis \etal 2000; Balbi \etal 2000).
Retrospectively, this model is close to
WMAP (Wilkinson Microwave Anisotropy Probe) 
normalized model (Spergel \etal 2003).
The simulation box size is $L_{box}=25h^{-1}$Mpc
having $768^3$ gas cells and $384^3$ dark matter particles
with the mean baryonic mass in a cell being 
$3.3\times 10^5h^{-1}\msun$ and
the dark matter particle mass being
$2.0\times 10^7h^{-1}\msun$.
The nominal spatial resolution for both gas and dark matter 
is comoving $32h^{-1}$kpc.
This mass resolution is considerably better than of most cosmological
simulations, but the spatial resolution, while significantly inferior to
that obtained in both the SPH and AMR schemes, is, we believe,
adequate for the present purpose.

%The assumed amplitude of the input power spectrum,
%as represented by $\sigma_8$ is nearly 20\%
%higher than the current best estimate for that quantity
%(Lahav \etal 2001).
%We expect that a computation based on the lower normalization
%would not alter any of the major conclusions
%reached in this study but would lower the absolute
%number of DLA systems at all redshifts,
%and we are currently initiating a new, still higher
%resolution simulation to test this expectation.

The initial conditions adopted are those 
for Gaussian processes with the phases
of the different waves being random and uncorrelated.
The initial condition is generated by the
COSMICS software package kindly provided by 
E. Bertschinger (2001).
The simulated box is an unconstrained and truly random realization 
of the model universe of that size.
Numerical methods of the cosmological hydrodynamic code 
and input physical ingredients 
have been described in our earlier papers
(Cen \& Ostriker 1999a,b).
Here we give a more detailed description.

The simulation integrates five sets of equations
simultaneously: 
the Euler equations for gas dynamics,
rate equations for different hydrogen and helium 
species at different ionizational states,
the Vlasov equation for dynamics of collisionless particles,
the Poisson's equation for obtaining the gravitational potential field
and the equation governing the evolution of the intergalactic
ionizing radiation field,
all in cosmological comoving coordinates.
The gasdynamical equations are solved using 
the TVD shock capturing code (Ryu \etal 1993) on an uniform mesh.
The rate equations are treated using sub-cycles within a hydrodynamic
time step due to much shorter ionization timescales
(i.e., the rate equations are very ``stiff").
Dark matter particles are advanced in time using the standard
particle-mesh (PM) scheme.
The gravitational potential on an uniform mesh is solved
using the Fast Fourier Transform (FFT) method.

The radiation field from $1$eV to $100$keV
is followed in detail
with allowance for
self-consistently produced radiating sources and sinks in the simulation box
and for cosmological effects, i.e., radiation transfer
for the mean field $J_\nu$ is computed with stellar,
quasar and bremsstrahlung sources and sinks due to \lya clouds etc.
In addition, a local optical depth approximation is adopted to crudely mimic
the local shielding effects: each cell is flagged with an
hydrogen ``optical depth" equal to the product of neutral hydrogen
density, hydrogen ionization cross section and the cell size;
equivalent ones for neutral helium and singly-ionized helium are 
also computed.
In computing the global sink terms for the radiation field
the contribution of each cell is subject to the shielding 
due to its own ``optical depth".
In addition, in computing the local ionization and cooling/heating
balance for each cell the same shielding is taken into account
to attenuate the external ioinizing radiation field.

The simulation computes for each cell and each timestep
detailed cooling and heating
processes due to all the principal line and continuum 
processes for a plasma of primordial composition.
Metals ejected from star formation (see below)
are followed in detail in a time-dependent, non-equilibrium fashion.
Cooling due to metals is computed 
using a code based on the Raymond-Smith code
assuming ionization equilibrium (Cen \etal 1995):
at each time step, given the
ionizing background radiation field,
we compute a lookup table for metal cooling 
in the temperature-density plane 
for a gas with solar metallicity,
then metal cooling rate for each gas cell 
is computed using the appropriate entry 
in that plane multiplied by its metallicity (in solar units).
In addition, Compton cooling due to the microwave background
radiation field and  Comption cooling/heating due to 
the self-consistently produced (see below)
X-ray and high energy background are also included.

We follow star formation using a well defined (heuristic but plausible)
prescription
used by us in our earlier work
(Cen \& Ostriker 1992,1993) and 
similar to that of other investigators 
(Katz, Hernquist, \& Weinberg 1992;
Katz, Weinberg, \& Hernquist 1996;
Steinmetz 1996;
Gnedin \& Ostriker 1997).
A stellar particle of mass
$m_{*}=c_{*} m_{\rm gas} \Delta t/t_{*}$ is created
(the same amount is removed from the gas mass in the cell),
if the gas in a cell at any time meets
the following three conditions simultaneously:
(i) flow contracting, (ii) cooling time less than dynamic time, and 
(iii) Jeans unstable,
where $\Delta t$ is the timestep, $t_{*}={\rm max}(t_{\rm dyn}, 10^7$yrs),
$t_{dyn}$ is the dynamical time of the cell,
$m_{\rm gas}$ is the baryonic gas mass in the cell and
$c_*=0.25$ is star formation efficiency.
In essence, we follow 
the classic work of Eggen, Lynden-Bell \& Sandage (1962)
and assume that the dynamical free-fall and galaxy formation timescales
are simply related.
Each stellar particle has a number of other attributes at birth, including 
formation time $t_i$, initial gas metallicity
and the free-fall time in the birth cell $t_{dyn}$.
The typical mass of a stellar particle in the simulation
is about one million solar masses;
in other words, these stellar particles are like
coeval globular clusters.

Stellar particles are subsequently treated dynamically
as collisionless particles,
except that feedback from star formation is allowed in
three forms: UV ionizing field, supernova kinetic energy, and metal rich gas,
all being proportional to the local star formation rate.
Supernova energy and metals  
are ejected into the local gas cells where
stellar particles are located.
Supernova energy feedback
into the intergalactic medium (IGM) is included 
with an efficiency (in terms of rest-mass energy of total formed stars) of
$e_{SN}=1\times 10^{-5}$; i.e.,
for $m_*$ of stars formed $e_{SN} m_* c^2$ amount of energy from
supernovae is released to the IGM, where $c$ is the speed of light.
Two types of ionizing radiation sources are used:
one characteristic of star formation regions and the other
characteristic of quasars, with efficiencies (i.e., the fraction of
rest-mass energy converted into ionizing radiation) of
$e_{UV,*}=3\times 10^{-6}$, and $e_{UV,Q}=5\times 10^{-6}$, respectively.
We adopt the emission spectrum of massive stars from Scalo (1986) 
and that of quasars from Edelson \& Malkan (1986).
In addition, hot, shocked regions (like clusters of galaxies)
emit ionizing photons due to bremsstrahlung radiation.
The temporal release of the all three feedback components at time $t$
has the same form:
$(dt/ t_{dyn}) [(t-t_i)/t_{dyn}]\exp[-(t-t_i)/t_{dyn}]$,
where $t_i$ is the formation time of a stellar particle.
The UV component is simply averaged over the box,
since the light propagation time across our box
is small compared to the timesteps.

It is useful to state clearly how we treat the chemical evolution in
our simulations.
We did not separately make any adjustments to fit to the observed distributions
and evolution of metals,
but assumed a specific efficiency of metal formation,
an adopted ``yield" (Arnett 1996), 
the percentage of stellar mass that is ejected back into IGM as metals,
of $0.02$.
This is the only adjustable parameter which significantly
effects metallicity and, to zero-th order, all quoted metallicity
scale with this assumed yield.
Metals in the IGM (assuming the standard solar composition)
are followed as a separate variable 
(analogous to the total gas density)
with the same hydrocode.
In essence, we compute in current simulations
the metals enrichment process of the IGM due to type II supernovae.
Furthermore, we use Zn to represent the total metals (i.e., O),
i.e., we adopt $[Zn/O]=0$.
While the origin of Zn may be unclear, 
our justification is purely empirical based on observations.
The assumption of $[Zn/O]=0$
is consistent with the extant, albeit,
small set of DLAs whose oxygen abundances have been detected
(Prochaska \& Wolfe 2002; Pettini \etal 2002);
there is no evidence to the contrary.
%since iron abundance at low metallicity such as in DLAs
%is primarily contributed by type II SNe. 
We convert observed metallicities of DLAs expressed in 
other species to Zn abundance,
calibrated by observed empirical relations for DLAs.

%Since the model has supernova input,
%the star forming galaxies 
%[equivalent to the observed Lyman break galaxies (LBGs)]
%can generate winds which pollute their surroundings with metal enriched
%gas.

The model reproduces the observed UV background as a function of redshift,
the redshift distribution of star formation
(``Madau Plot"; Nagamine, Cen \& Ostriker 1999),
and 
the galaxy luminosity function (Nagamine \etal 2001b).
We find that the computed metallicity distributions
over a wide range of environments, including 
clusters of galaxies, damped Lyman systems, Lyman alpha forest
and stars, are in broad agreement with observations
(Cen \& Ostriker 1999b; Nagamine \etal 2001a,b; Cen \etal 2001),
lending us confidence that the computed metal distribution
in the intermediate regions 
under consideration 
(between Lyman alpha forest and clusters of galaxies)
in this high resolution simulation
may be a good approximation to the model universe.

We identify a DLA with 
a system with a line-of-sight (LOS) neutral hydrogen column density 
higher than $2\times 10^{20}$cm$^{-2}$.
This method is similar to what 
is employed to observationally classify DLAs.
In addition, we can easily identify, in three-dimension,
host galaxies that are responsible for DLAs.
A host galaxy is associated with a DLA 
by comparing their spatial locations.
Galaxies in the simulation box are identified using a scheme
described in Nagamine \etal (2000), with each having
a set of attributes including mean formation redshift and
initial gas metallicity.
Each galaxy consists of sub-units each with formation time,
mass, metallicity and we compute the luminosity
of each galaxy using a population synthesis code (e.g., Bruzual 2000).
Each galaxy is also associated with the dark matter mass
around it or dark matter halo and virial velocity of the whole system.
We note that multiple DLAs may be associated with a single
host galaxies, simply because different LOS
through a galaxy may be observed as DLAs.
The typical impact parameter is comparable to 
the cell size.
Throughout the paper we refer to "DLA galaxies" as
the host galaxies of DLAs.

\section{Results}

\subsection{Pictures}

\begin{figure}
\plotone{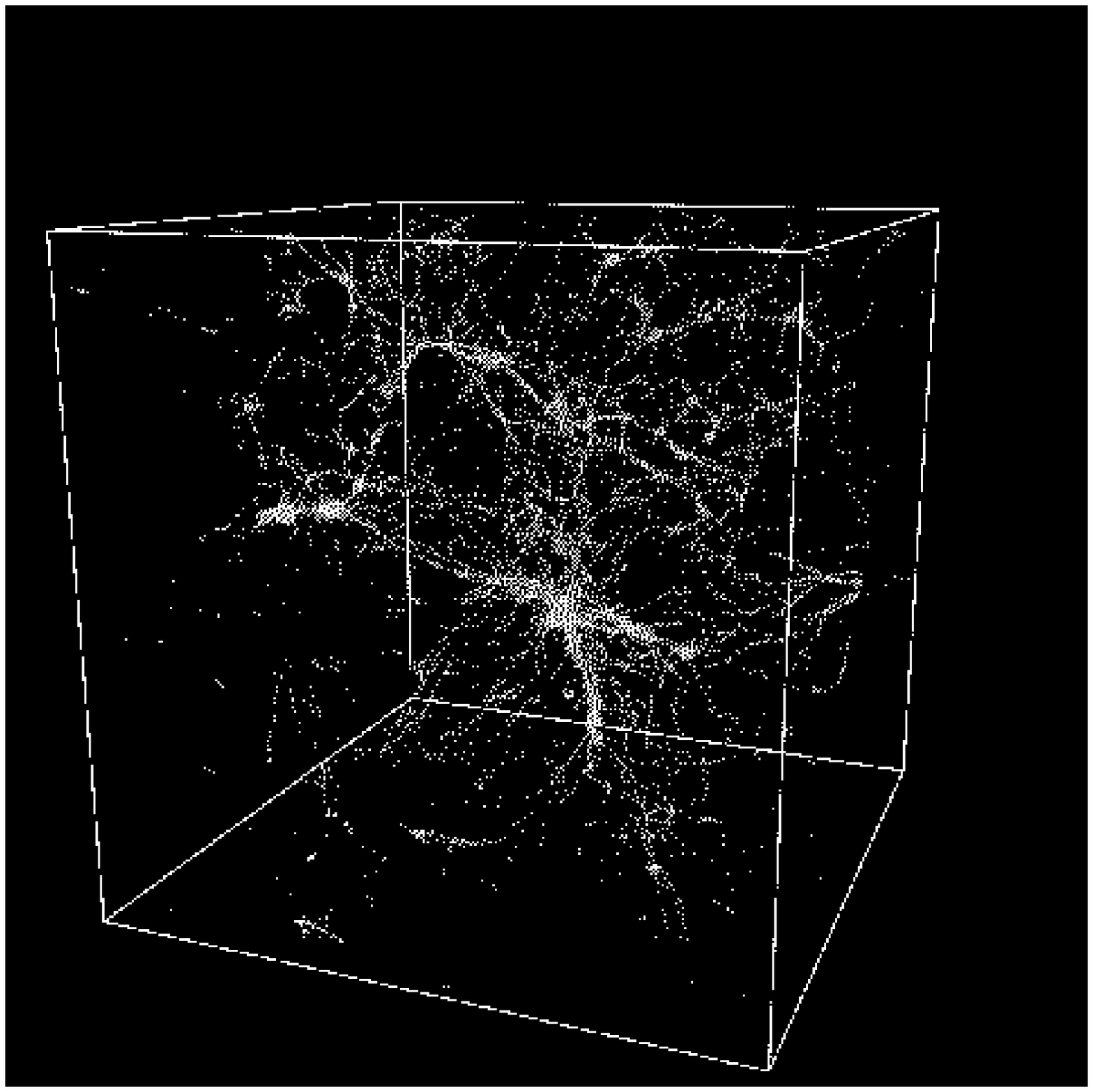}
\caption{
shows the total gas phase at $z=3$ color coded. 
The box size is $25h^{-1}$Mp.
The green regions have an overdensity of order $10$,
while the yellow regions have 
an overdensity of order $100$.
Damped Lyman alpha systems are typically located
in the highest density  regions denoted in red,
which have overdensity of order $1000$ or higher.
}
\label{fig1}
\end{figure}

\begin{figure}
\plottwo{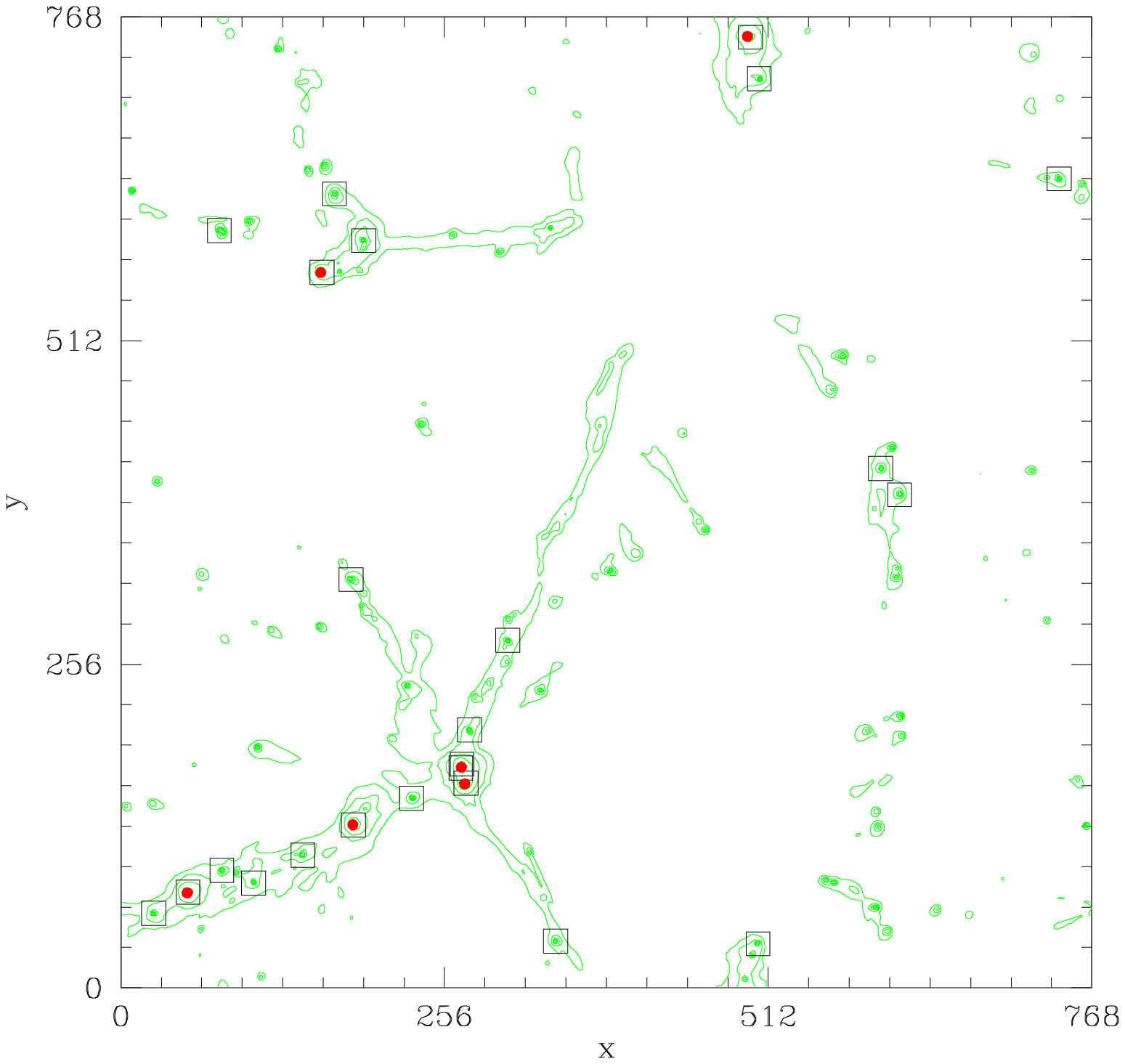}{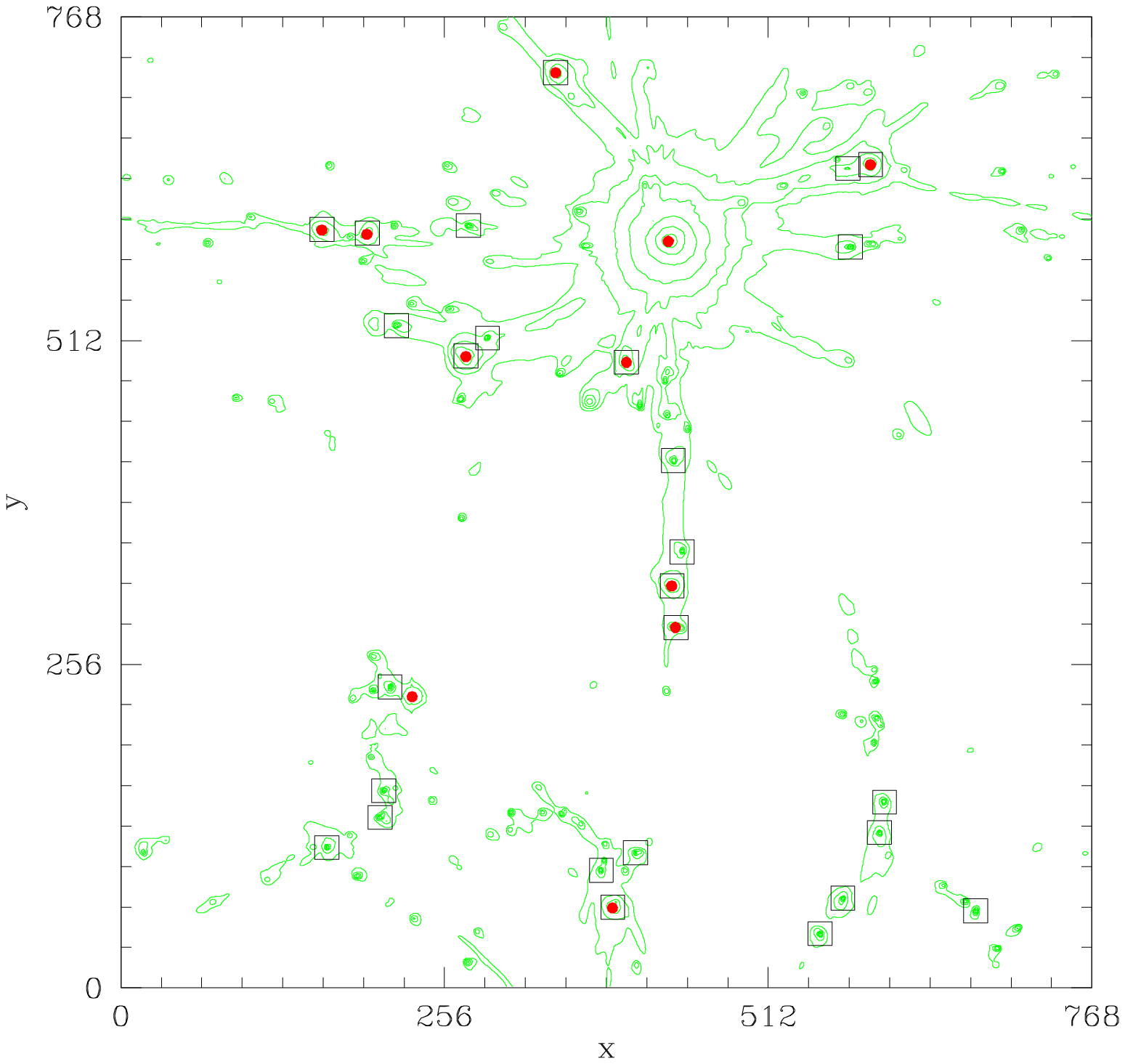}
\caption{
shows two slices of gas density distribution
with a size of $25\times 25 \times 3.125h^{-3}$Mpc$^3$ each 
at $z=0$.
The left slice is a typical one and 
the left slice contains the richest cluster in the simulation.
The green contours show the total gas density.  
The outermost contour has the mean density 
and each successive contour has a density higher by 0.5 dex.
Regions that contain DLAs in the same slices are indicated by  black 
squares.
$L^*$ galaxies are denoted by red dots.
Note that DLA systems are in regions of moderate over-density
avoiding the highest
density, highest temperature regions.
}
\label{fig2}
\end{figure}

A visual impression of the gas distribution in our
box is presented in Figure (1), which
shows the total gas distribution at redshift $z=3$.
DLAs are typically located
in or near the highest density  regions denoted in red.
More quantitatively, 
in Figure (2) we show two slices of gas density distribution
with a size of $25\times 25 \times 3.125h^{-3}$Mpc$^3$ each 
at $z=0$.
The left slice is typical and 
the right slice contains the richest cluster in the simulation.
We see that rich clusters are hostile environments for DLAs; 
DLAs tend to reside in filaments which bridge
and surround clusters.
This is due to the same physical effect described
in Blanton \etal (1999).
DLAs are positively 
correlated with high total density but negatively correlated
with high environmental temperature (at least at the high temperature end);
the specification of one variable - density -
does not enable one to predict the likely
occurance of collapsed systems.
While more quantitative comparisons can not be made,
this finding appears to be consistent with 
the fact that the observed spiral galaxies in clusters have
lower HI content (Haynes \etal 1990; Haynes \& Giovanelli 1991).
In Figure 2 red dots show galaxies having stellar mass $\ge 10^{10.5}\msun$
and we see that roughly 30\% of DLAs today are associated
with massive galaxies like the Milky Way.
We will return to this issue to present
more quantitative results later.

\subsection{Quantitative Results}

\begin{figure}
\plotone{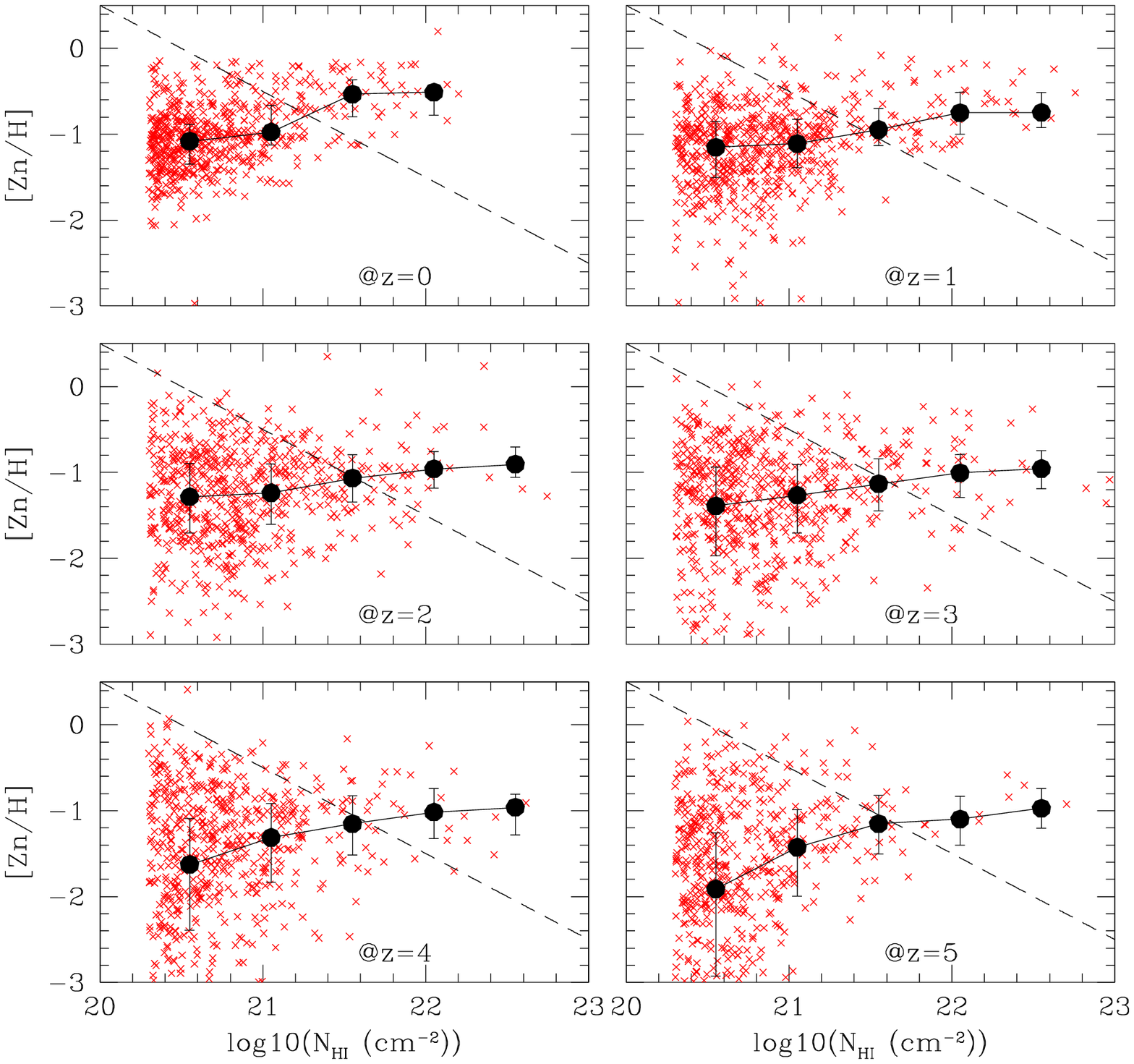}
\caption{
shows metallicity distribution of DLAs
as a function of neutral hydrogen column density.
Each ``x" represents a line-of-sight DLA
and the large solid dots represents the median
metallicities with quartile errorbars for a few column density bins.
In each panel
a dashed line with $\log10(N_{HI})+ [Zn/H] = 20.5$ is included.
}
\label{fig3}
\end{figure}

\begin{figure}
\plotone{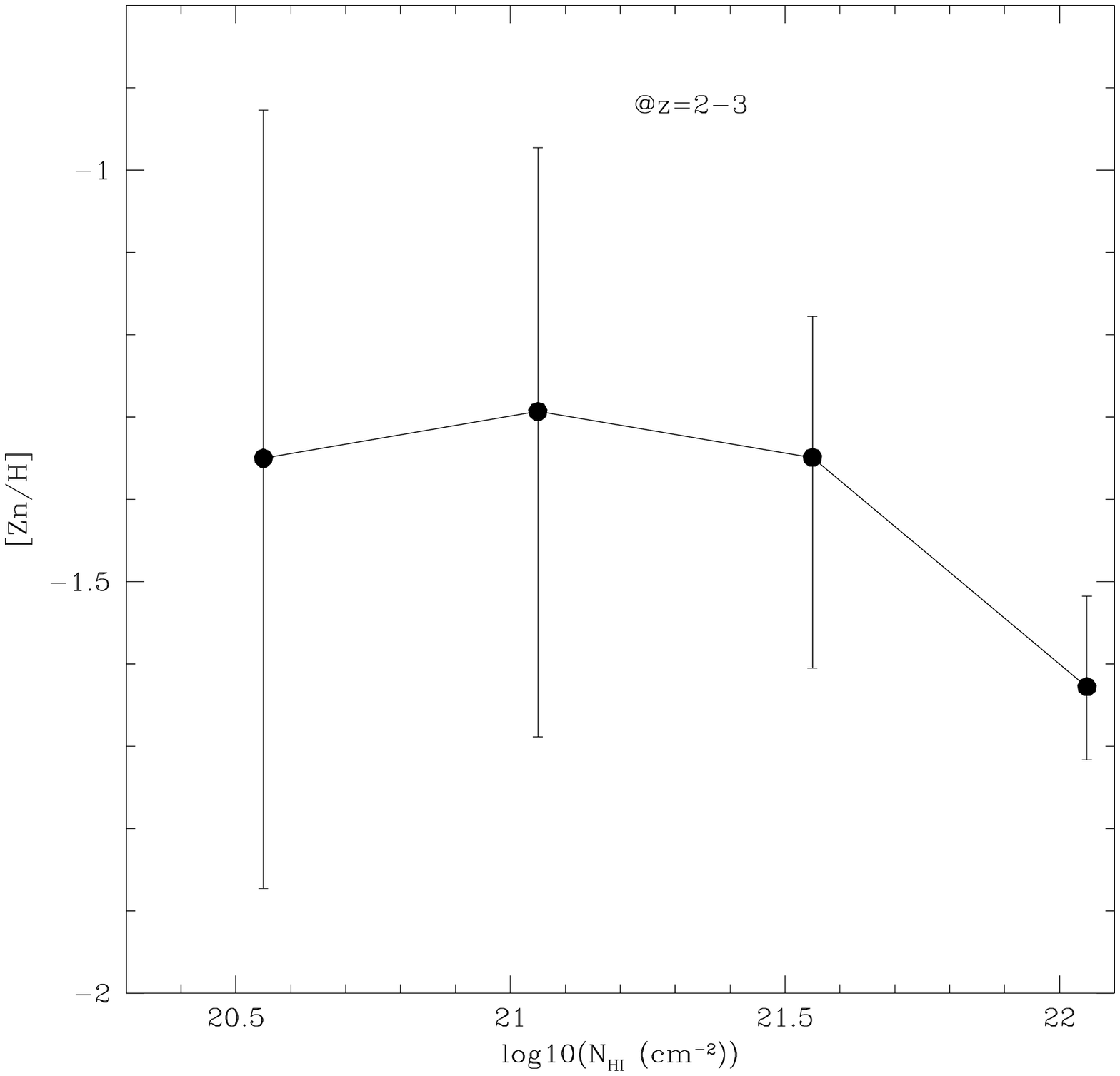}
\caption{
shows with the median metallicities with quartile errorbars 
as a function of column density for simulated DLAs at 
$z=2-3$.
}
\label{fig4}
\end{figure}

Let us now turn to quantitative results.
Figure (3) shows the metallicity distribution of DLAs
as a function of neutral hydrogen column density, $N_{HI}$.
At each redshift there is a weak trend of higher metallicity
correlating with higher column density.
Although the variance is large, as indicated by the quartile errorbars,
this is, at first sight, 
inconsistent with observations where the opposite trend is seen.
There may be important observational selection effects that
have not yet been included in our analyses,
which have caused this apparent inconsistency.
In particular, dust obscuration may have played a large role.
We will explore this possibility here.

To illustrate the dust obscuration effect we show, in each panel
in Figure 3,
a dashed line with $\log(N_{HI})+[Zn/H] = 20.5$, %- log(1+z)$],
intending to approximately separate the region
that is expected to be affected by dust obscuration (upper right)
and the region that is not (lower left),
partly motivated by observations. %(Boiss\'e \etal 1998).
If DLAs in the upper right side of the dashed lines had obscured QSOs,
one would expect to see a large deficit of DLAs there.
In fact, 
there is some observational indication
that this effect may be there (Boiss\'e \etal 1998).
In Figure 4 we plot  %500 DLAs randomly chosen 
                     %from a total sample of 14287 DLAs
%obtained by simply summing up all DLAs found 
%in the simulation box at four epochs, $z=0,1,2,3$,
%that lie below the dashed lines in Figure 3a.
%obtained using all the 14287 DLAs.
the median metallicities (with quartile errorbars) 
for the ``unobscured" DLA population at $z=2-3$.
We see that 
the weak positive correlation between the median
metallicity and column density of DLAs seen in Figure 3
is now reversed, with a result that is consistent 
with the observed trend (Boiss\'e \etal 1998). % given the large errorbars.

\begin{figure}
\plotone{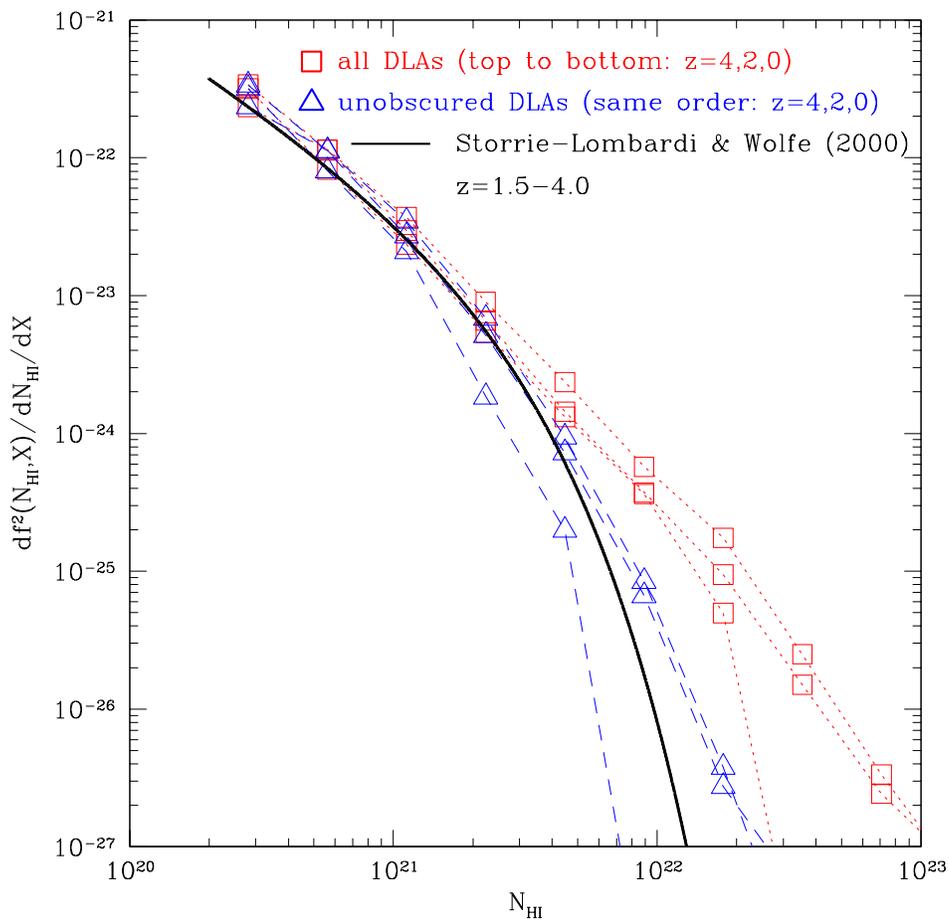}
\caption{
shows column density distributions of 
all DLAs (squares) 
and ``unobscured" DLAs (triangles)
at three redshifts, $z=4,2,0$, 
from top to bottom.
The ordinate is the number of DLAs
per unit column density per unit ``absorption distance",
defined in Bahcall \& Peebles (1969).
Also shown as a think solid curve is the Schechter fit of
the observed DLA distribution according to
Storrie-Lombardi \& Wolfe (2000) for the redshift range $z=1.5-4.0$.
It is seen that the computed DLA column density distribution
agrees well with observations.
}
\label{fig5}
\end{figure}

As a consistency check we show in Figure 5 the column density distributions
of DLAs at three redshifts ($z=0,2,4$).
We plotted the total DLA population (squares)
as well as the ``unobscured" DLA population, as defined 
in Figure 3.
Quite surprisingly, the ``unobscured" DLA population 
provides a remarkably good fit to observations 
shown as the thick solid curve (Storrie-Lombardi \& Wolfe 2000).
Therefore, our assumption of dust obscuration is at least self-consistent
and the model would provide a reasonable fit to observations,
if the observed DLAs are a subset of total DLA population
that is not obscured by dust.
There is some indication 
that the model may be producing too many DLA
at the low column density end even
with the dust obscuration effect taken into account.
This, however, may be telling us something
about the small-scale power in the adopted cosmological model.
It may be explained if 
the adopted LCDM model has too much small
scale power, resulting in more numerous dense neutral cores; 
note that the small-scale power makes a much 
larger contribution to the DLA population at high redshift.
This implication would be consistent with other independent
observations including dwarf galaxies
(Flores \& Primack 1994;
Burkert 1995;
Klypin \etal 1999; 
Moore \etal 1999;
McGaugh \& de Blok 1998;
Navarro \& Steinmetz 2000).
However, it could also be due to limited numerical simulation
of the present simulation, which artificially enlarges
density concentrations and thus produces more DLAs than
there should be.
Future high resolution simulations can check this.
However, given that the metallicity dependence on
column density is very weak, the conclusions drawn
with regard to the metallicity of DLAs and its evolution
are relatively robust.

Visual examination of Figure 3 may give the reader the
impression that there appears to exist a significant
number of DLAs at $N_{HI}>10^{21}$cm$^{-2}$
in the model even for the ``unobscured"
DLA sub population.
A more quantitative comparison with observations 
is now made.
The simulation and observations indicate, respectively, 
that about (30\%, 27\%) of all DLAs have $N_{HI}>10^{21}$cm$^{-2}$ at $z=2$.
At $z=3$ simulation produces
about 26\% of DLAs having $N_{HI}>10^{21}$cm$^{-2}$ ,
while the corresponding observed number is 23\%.
The simulation is also in good agreement with
observations at low redshift ($z\sim 0-0.5$) where a high rate of occurance
of such high column density DLAs is seen 
(Rao \& Turnshek 2000).
These results are consistent with the close agreement seen in Figure 5.
At $z>3.5$, however,
simulation gives 20\% of DLA with $N_{HI}>10^{21}$cm$^{-2}$
at $z=4$ versus 2.5\% for observed DLAs with $N_{HI}>10^{21}$cm$^{-2}$.
While the number of observed DLAs at the highest redshift range
is still small, this discrepancy at high redshift
may be revealing the same alleged flaw of excessive small-scale power
in the simulated model. 

\begin{figure}
\plotone{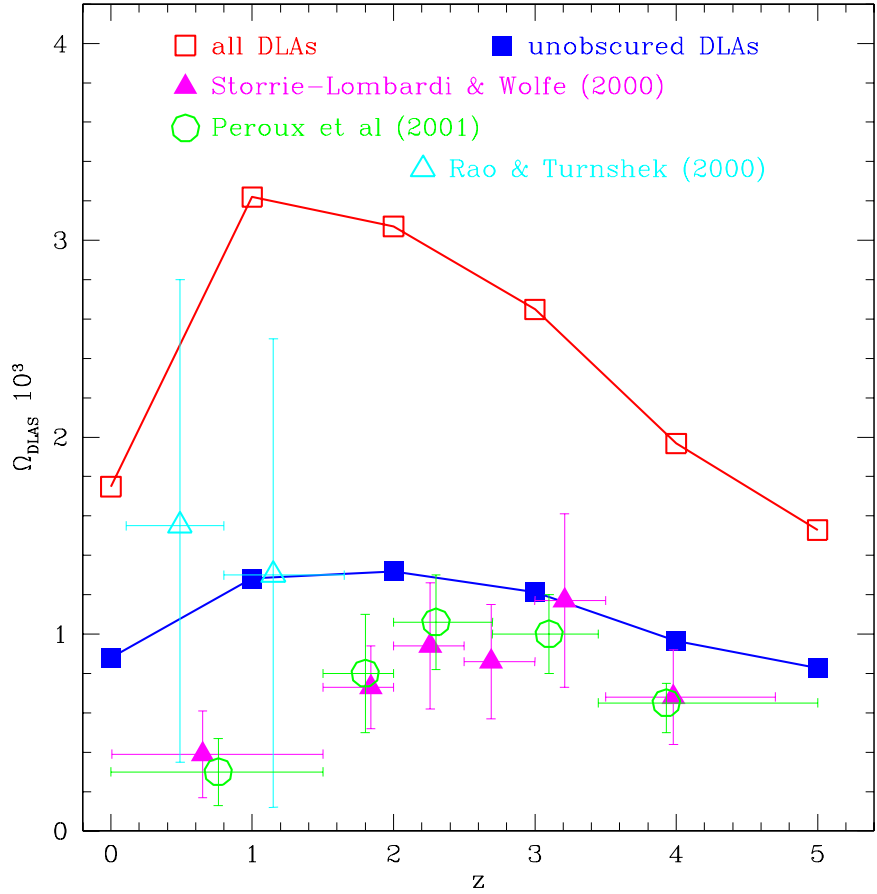}
\caption{
shows total neutral gas contained in DLAs (open squares)
and neutral gas in ``unobscured" DLAs (filled squares)
as a function of redshift.
Shown as symbols
are observations from
Storrie-Lombardi \& Wolfe (2000; solid triangles),
Peroux \etal (2001; open circles)
and Rao \& Turnshek (2000; open triangles).
}
\label{fig6}
\end{figure}

The neutral gas contained in DLAs 
is another extensively studied quantity,
for its obvious link to stars in that the latter
must form out of the former,
and is closely related to the column density distribution shown
in Figure 5.
(But note that the converse statement is not necessarily
true; not all DLAs are transformed to galaxies.
We find many fall into clusters of galaxies and are added to hot, shocked gas.)
It thus seems warranted to present 
the results from the simulation.
The reader is reminded that, unlike the metallicity distributions
but like the column density distribution shown in Figure 5,
the computed fraction of the neutral gas in DLAs 
is much more prone to numerical effects as well as variations in the
cosmological model.
Figure 6 presents  
the neutral gas in DLAs as a function of redshift,
compared to extant observations.
Note that both the computed and observed
neutral gas in DLAs are subject to the uncertainties
on the upper limit of the column density distribution,
since it is dominated the highest column density DLAs 
(given that the slope of the column density distribution
is substantially flatter than $-2.0$).
Consequently but unsurprisingly,
the computed neutral gas in total DLAs is 
significantly (about a factor $1.5-2.5$) higher than
the observed values at all redshifts,
in agreement with the overabundance of high column density
DLAs in the simulation seen in Figure 5.
But the ``unobscured" DLAs provide a reasonable
fit to observations, given both observational
errors (especially at low redshifts)
and numerical uncertainties.
Our assumption of dust obscuration is again self-consistent.

Clearly, the real situation due to dust obscuration 
is much more complicated (see Fall \& Pei 1993 for a more detailed
treatment). 
For example,
it is not clear what the relationship between dust content
and metallicity should be and how large the scatter is.
In addition, dust properties are largely uncertain.
Nevertheless, such an approximate treatment 
evidently does have a quite large effect.
It appears that
the computed model would be in fair agreement with 
observations with regard to several major diagnostics shown above,
if dust obscuration has played an important role. 
It is, however, too early to celebrate the existence of dust obscuration.
There are other observations that indicate
dust obscuration effect may not be as large as 
considered here (Ellison \etal 2001; Prochaska \& Wolfe 2002).
It is worth noting that we find that in the redshift
range $z=2-3$ the fraction of the number of 
excluded DLAs due to dust obscuration (those that lie above
the dashed lines in Figure 3)
is 9.9\%.
Therefore, in the CORALS radio survey of Ellison \etal (2001)
one would have expected to see, on average, only $1.8$ DLAs, that
are dust obscured, in their sample of total $19$ DLAs.
Clearly, a sample size of several-fold larger than the current one
would greatly firm up the statistical significance of 
their important findings and determine the importance of dust
obscuration unambiguously.
Although the main conclusion of this paper that the metallicity 
of DLAs is substantially sub-solar and 
its evolution is relatively mild 
does not require the adoption of dust obscuration as proposed,
other properties of DLAs, including column density distribution,
evolution of neutral gas content 
and metal content in DLAs (see Figure 17 below
and related text),
are all seen to become consistent with observations under that assumption.
Alternatively,
there are at least two other possibilities to potentially explain
the apparent discrepancies.
First, the adopted
cosmological model may contain flaws and require revision with regard
to its properties related to the formation of DLAs.
However, the adopted model is known to reproduce many other observations
on scales that overlap with that related to DLAs, including
galaxy abundance and clustering,
although there is independent evidence that the adopted
model may possess excessive power on small scales (see below).
Second, it is possible that astrophysical processes
related to star formation have not been treated adequately.
In particular, metal enrichment of the intergalactic medium treated
here is clearly oversimplified. We think this is a possibility
that requires substantially more work and 
deserves further investigation with future simulations that
possess more refined treatments of galaxy formation.

\begin{figure}
\plotone{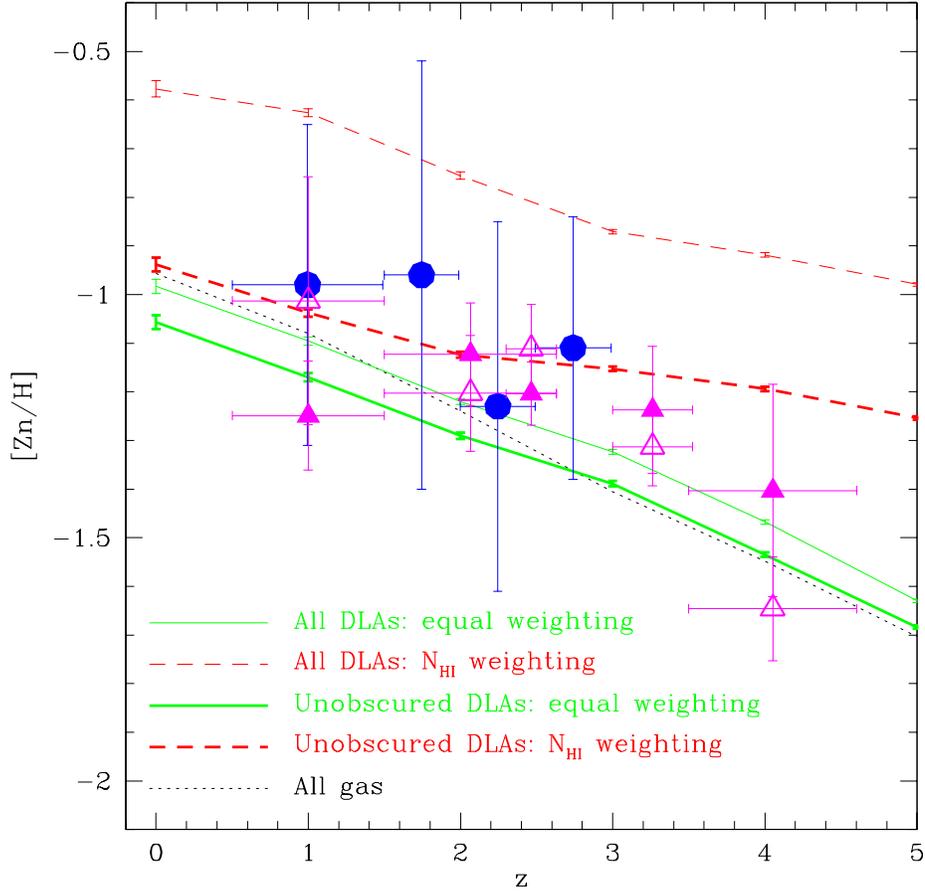}
\caption{
shows unweighted (thin solid curve) 
and neutral hydrogen column weighted (thin dashed curve)
metallicity 
averaged over all DLAs identified in the simulation
as a function of redshift.
The thick solid and thick dashed dotted curves 
are the equivalent
averaged over only the DLAs located below the lines Figure 3.
Observations are shown as symbols
where solid dots are 
column density weighted mean from Pettini \etal (1999),
and solid and open triangles  
column density weighted and equally weighted means
from Prochaska \& Wolfe (2000).
All observed metallicity with 
[Fe/H] 
is corrected to 
[Zn/H] adopting [Fe/Zn]=-0.4 (Prochaska \& Wolfe 2002).
Also shown as a thin dotted curve
is the metallicity averaged over all the gas in the universe.
}
\label{fig7}
\end{figure}

Let us now return to the main 
issue of metallicity evolution of DLAs.
To compute this,
we adopt two averaging methods used in analyzing observational data sets.
The first method weights each DLA equally,
$<[Zn/H]>_{1}\equiv {\sum_{N} [Zn/H]\over N}$,
and the second method weights each DLA
by its neutral hydrogen column density,
$<[Zn/H]>_{2}\equiv \log_{10}\left({\sum N_{Zn} \over \sum N_{HI}}\right)$.
In Figure 7 we show 
$<[Zn/H]>_{1}$ (thin solid curve) 
and 
$<[Zn/H]>_{2}$ (thin dashed curve) as a function of redshift,
averaged over all DLAs identified in the simulation.
The thick solid and thick dashed dotted curves in Figure 7
are 
$<[Zn/H]>_{1}$ and
$<[Zn/H]>_{2}$ 
averaged over only the ``unobscured" DLAs.
Also shown in Figure 7 as a thin dotted curve
is the metallicity averaged over all the gas in the universe
(using the second averaging method).

Several interesting points may be extracted from Figure 7.
The evolution of $<[Zn/H]>_{1}$ of all DLAs (thin solid curve)
follows rather closely 
the universal metallicity evolution of all gas (thin dotted curve),
which is mainly a coincidence.
The evolution of $<[Zn/H]>_{2}$ of all DLAs (thin dashed curve),
while approximately parallel to 
$<[Zn/H]>_{1}$ and evolution of all gas,
has an amplitude about 0.5 dex higher.
This indicates that formation of dense neutral cores occurs
preferentially in regions that have higher metallicity
than the average regions of the universe at all epochs,
with the trend that deviation from the mean increases with redshift.
However, this systematic difference between 
$<[Zn/H]>_{1}$ and $<[Zn/H]>_{2}$ is substantially reduced to $\sim 0.2$ dex,
if dust obscuration has a significant selection effect on DLAs,
as evident from a comparison between the two thick curves,
except at the high redshift end ($z\ge 3$).

There appears to be a reasonable agreement between
the model and observations 
with regard to the evolution of 
metallicity of unobscured DLAs 
for both neutral hydrogen column density weighted 
and unweighted means.
There is some mild indication that 
the unweighted mean of the model may be a little lower than
observed or equivalently, there is an overabundance
of low metallicity low column density DLAs in the simulation.
A possible explanation may be that the current simulation
underestimates star formation in small galaxies due to limited resolution;
with an ideal resolution, 
metal enrichment from additional stars in small galaxies 
would have raised the metallicity in very low metallicity 
regions ($[Zn/H]\le -3$) in the present simulation to a higher level.
Given the cost of the simulation, we will have to investigate
this issue in future simulations.
In addition, our oversimplified treatment of chemical evolution
might have contributed to the discrepancy.

The most striking property that
is borne out from both the simulation and observations
is that a typical DLA has a metallicity substantially below the solar value
and does not show significant evolution with redshift.
Both can %easily
be understood by considering selection bias and
the source and sink terms for objects termed DLAs.
If one were to follow, in a Lagrangian sense, a typical
region of dense gas one would find it steadily
merging with other such units and
finding itself
(on average) in systems of higher and higher observed neutral
hydrogen column density.
Also, since star formation, whether steady or episodic,
leads to a monotonic increase in metallicity,
the fluid element would show an ever higher metallicity if observed.
At some point in time it would be in a system with neutral
column density $N_{HI}>10^{20}$cm$^{-2}$ and,
when observed would be called
a ``DLA" system.
The physical system in which it is embedded would evolve
to higher $[Zn/H]$ over time and ultimately
be identified with either an optical galaxy or,
if it should find itself in a very high temperature high
density region and the gas be stripped/evaporated,
to become an E or S0 system.
Typically by the time the metallicity had reached solar values
either gas would have been depleted (turned into stars),
expelled (by supernovae), or stripped due to interaction
with a hot phase intergalactic medium and so the system
would no longer have $N_{HI}>10^{20}$cm$^{-2}$.
All the processes are included in our code, but of course
limited by our numerical resolution.
Subsequent fluid elements passing along this same sequence
would have roughly the same metallicity $[Zn/H]$ at the same values
of column density $N$. Thus the $[Zn/H](N_{HI})$ relation would end up
evolving far more slowly as a function of redshift than would the
metallicity of a given fluid element.

\begin{figure}
\plotone{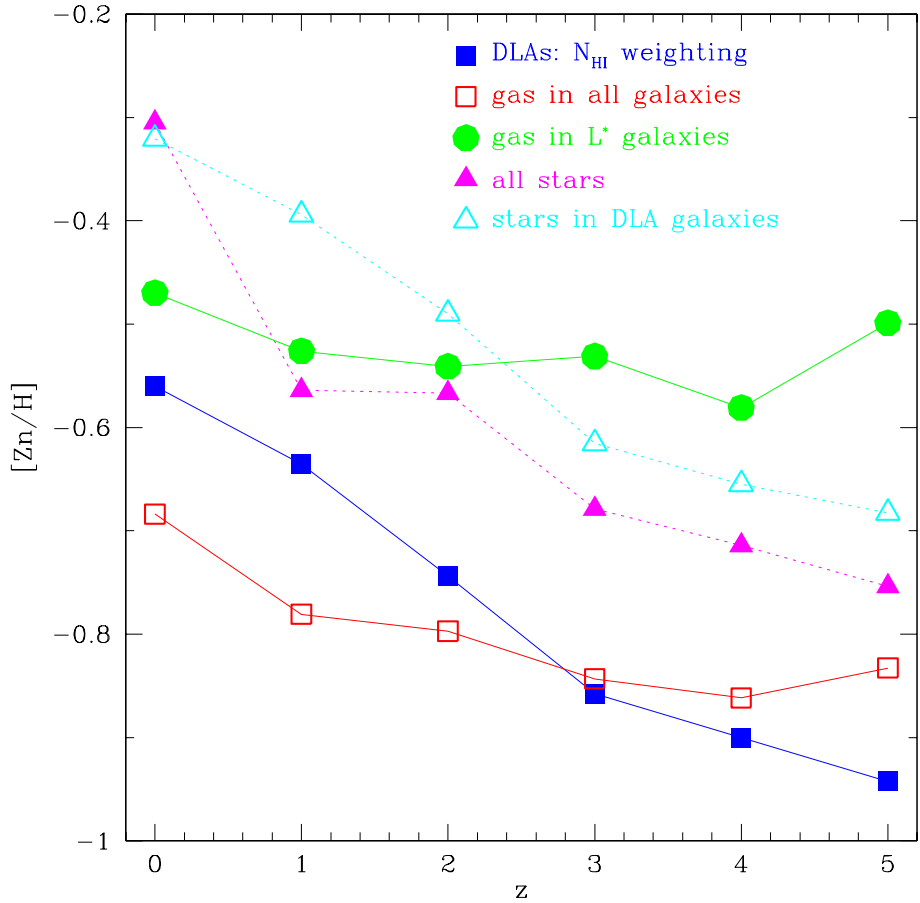}
\caption{
shows the metallicity of DLA gas,
gas in all galaxies and in massive ($\ge 10^{10.5}\msun$) galaxies,
all stars, stars in DLA galaxies and stars in $L^*$ galaxies.
Note $L^*$ is not constant in time, but increases as the typical
luminous galaxy becomes brighter.
}
\label{fig6}
\end{figure}

To further elucidate this physical picture,
we show, in Figure 8, a comparison between 
the metallicity of DLA gas and the metallicity of stars.
In addition to the globally averaged metallicity of DLA gas
we show the stellar metallicity of all stars and stars in DLA galaxies,
and gas in all galaxies and in massive 
(with stellar mass $\ge 10^{10.5}\msun$) galaxies,
respectively.
A few interesting features may be noted from Figure 8.
First, at all epochs the average stellar metallicity
is always higher than the gas metallicity of DLAs or all galaxies,
which in turn is higher than the average gas in the universe (see Figure 7).
This finding is very interesting and perhaps seems counter-intuitive
at first instance:
since stars are formed out of
gas in the galaxy, one naively would expect
that the gas should always have a higher metallicity than stars 
due to enrichment by subsequently formed stars.
This occurs in the ``closed box" model,
where the metallicity of gas in galaxies is typically
twice the metallicity of the stars in that galaxy.
But, in the hierarchical structure formation model,
fresh, lower metallicity gas 
is continuously accreted onto DLAs (or DLAs-to-be)
or DLAs form from mergers of smaller systems with gas of lower metallicity.
In addition, metal rich gas is steadily ejected from galaxies
through the action of supernovae.
This process has been invoked to solve the so-called G-dwarf problem 
(Ostriker \& Thuan 1975).
It is also interesting to note that 
the gas in the massive galaxies at high redshift ($z>2.5$) 
does have higher metallicities than their
resident stars,
probably indicating that
these massive galaxies constitute relatively closed boxes.
%because the high mass galaxies eject relatively little gas
%(Dekel \& Silk 1986; Mac Low \& Ferrara 1999).

Second, the gas in the most massive galaxies
has noticeably higher metallicity than both gas in DLAs and in all galaxies.
This is consistent with the picture outlined above in that
higher metallicity systems can not maintain the required
high neutral hydrogen column density due to either hotter
environment or hotter internal gas.

Third, stars in DLAs (open triangles) generally have higher metallicities
than typical stars (solid triangles), especially at redshift $z>1$.
This would be consistent with observations that less massive galaxies
generally have lower metallicity than higher mass galaxies,
if DLAs are more massive than a typical galaxy at high redshift.
This indeed turns out to be the case [see Figure 9 below,
where we show that DLAs are more massive than
the $L^*_{Sch}$ (one of the two parameters in the Schechter function)
galaxies at $z>2$.]
At lower redshift, the trend is somewhat reversed (also consistent
with Figure 9.)

Fourth, the metallicity of the gas in DLAs (solid squares)
displays
an interesting crossover at $z\sim 3$:
at $z>3$ the average gas
in all galaxies (open squares)
has a higher metallicity than the average gas in DLAs,
while the reverse is true at $z<3$.
This may indicate that at $z<3$
star formation in low mass systems 
is less efficient than in more massive systems 
so that gas is less metal enriched.
This trend is further consistent with the result that 
the gas in the most massive galaxies (solid dots in Figure 6)
has a metallicity higher than DLAs (solid squares).

Finally, we see by $z=0$ that
the gas even in the most massive galaxies 
has a metallicity lower than average stars or stars in DLAs
(stars in the same massive galaxies have comparable 
and very slightly higher metallicity than stars in DLAs at $z=0$;
not plotted).
If one associates the most massive galaxies with large spiral galaxies 
such as the Milky Way, one would expect to see gas metallicity
typically higher than stellar metallicity.
What is happening here is that these most massive galaxies
are elliptical systems
and live in clusters where galactic gas has been stripped off;
thus the gas metallicity for these massive galaxies is
actually the gas metallicity of the intra-cluster gas,
whose value ($1/3$ solar) seems consistent with observed
cluster gas metallicity 
(Arnaud \etal 1994; Mushotzky \etal 1996;
Tamura \etal 1996; Mushotzky \& Lowenstein 1997),
indicating that the adopted yield based on the theory
of stellar interior is approximately correct empirically.
Note that we do not include metal enrichment due to type Ia supernovae
in the present simulations.
%as a result, the agreement
%between the computed metallicity and the observed (mostly) iron
%abundance in the intracluster gas only serves as an rough indicator.
Just as a consistency check, we find that 
115 galaxies having [Zn/H]=[-0.1, 0.2],
which is quite consistent with the number of $L^*$ galaxies
that one expects from a simulation box of this size,
indicating that Milky Way-like galaxies can
properly be accounted for.

\begin{figure}
\plotone{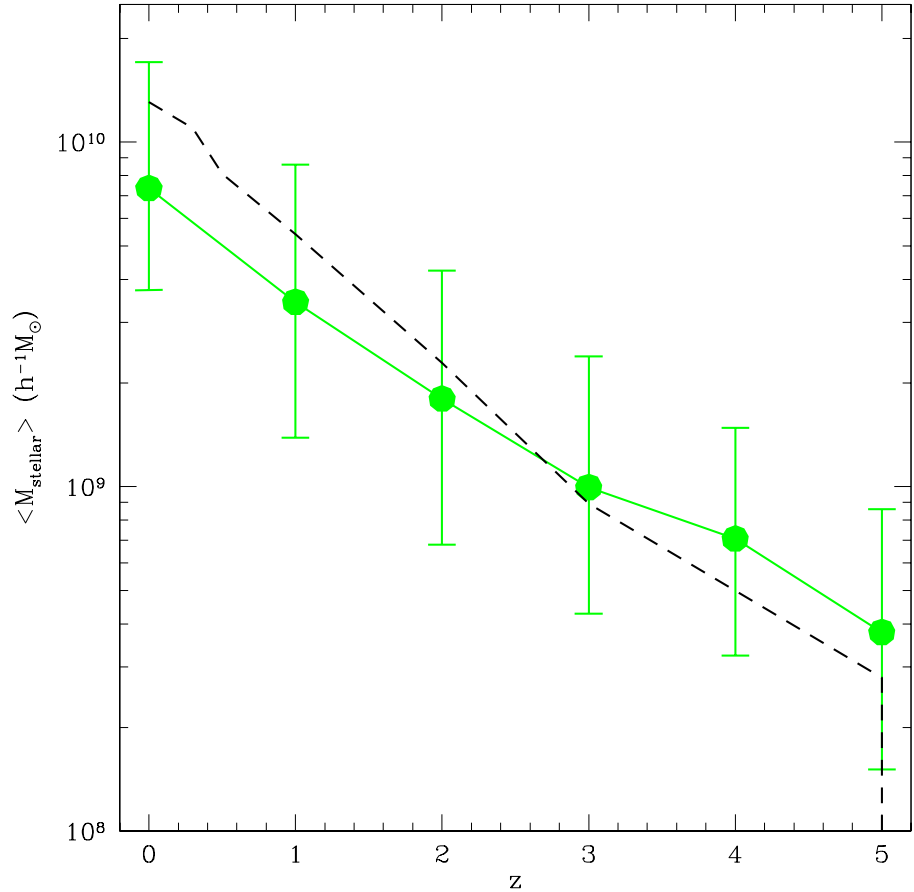}
\caption{
shows the median stellar mass (with quartile errorbars) of 
DLAs as a function of redshift.
Shown as dashed curve is the $M_*(z)$ (a parameter in 
the Schechter function fit; 
scaled using the parameters presented in Nagmaine \etal 2001).
}
\label{fig9}
\end{figure}

Needless to say galaxy formation and DLA formation are intimately related.
It is important to know what type of galaxies constitute
the host galaxies of DLAs.
In Figure 9 we show the median stellar mass (with quartile errorbars)
of DLAs as a function of redshift.
Not surprisingly, and consistent with the depicted description above,
DLAs correspond to progressively larger galaxies
towards lower redshift.
A typical DLA follows closely the $M_*$ at all redshifts,
with a slight tilt such that at low redshift ($z<2$)
a typical DLA is sub $M_*$ and the reverse is true at high
redshift ($z>2$).
While about 30\% 
of DLAs at $z=0$ are
due to $L^*$ galaxies (consistent with the visual inspection of Figure 2),
consistent with local observations (Rao \& Turnshek 2000),
DLAs at $z=3-5$ are typically in galaxies less luminous
than $0.1L^*(z=0)$ with some fractian 
in very faint galaxies ($L< 0.01L_*$).
The median luminosity of a DLA, $L_{DLA}(z)$,
in units of typical galaxy luminosity  at that redshift,
$L^*(z)$, that is, $(L_{DLA}/L^*)_z$ decreases
from $1.1$ to $0.5$ as redshift declines from $z=3$ to $z=0$,
but the absolute luminosity of the median DLA system
increases in the same interval by a
factor of five from 
$0.1L^*(z=0)$
to 
$0.5L^*(z=0)$.
We note that the stellar masses
for a typical DLA galaxy at 
$z=3$ is $M_{stellar} \sim 10^9\msun$,
about one order of magnitude lower than the stellar
masses deduced for LBGs (Shapley et al. 2001). 
This implies that DLAs and LBGs are different populations
or at least that a typical DLA galaxy does not correspond to an LBG
(Colbert \& Malkan 2001).

\begin{figure}
\plotone{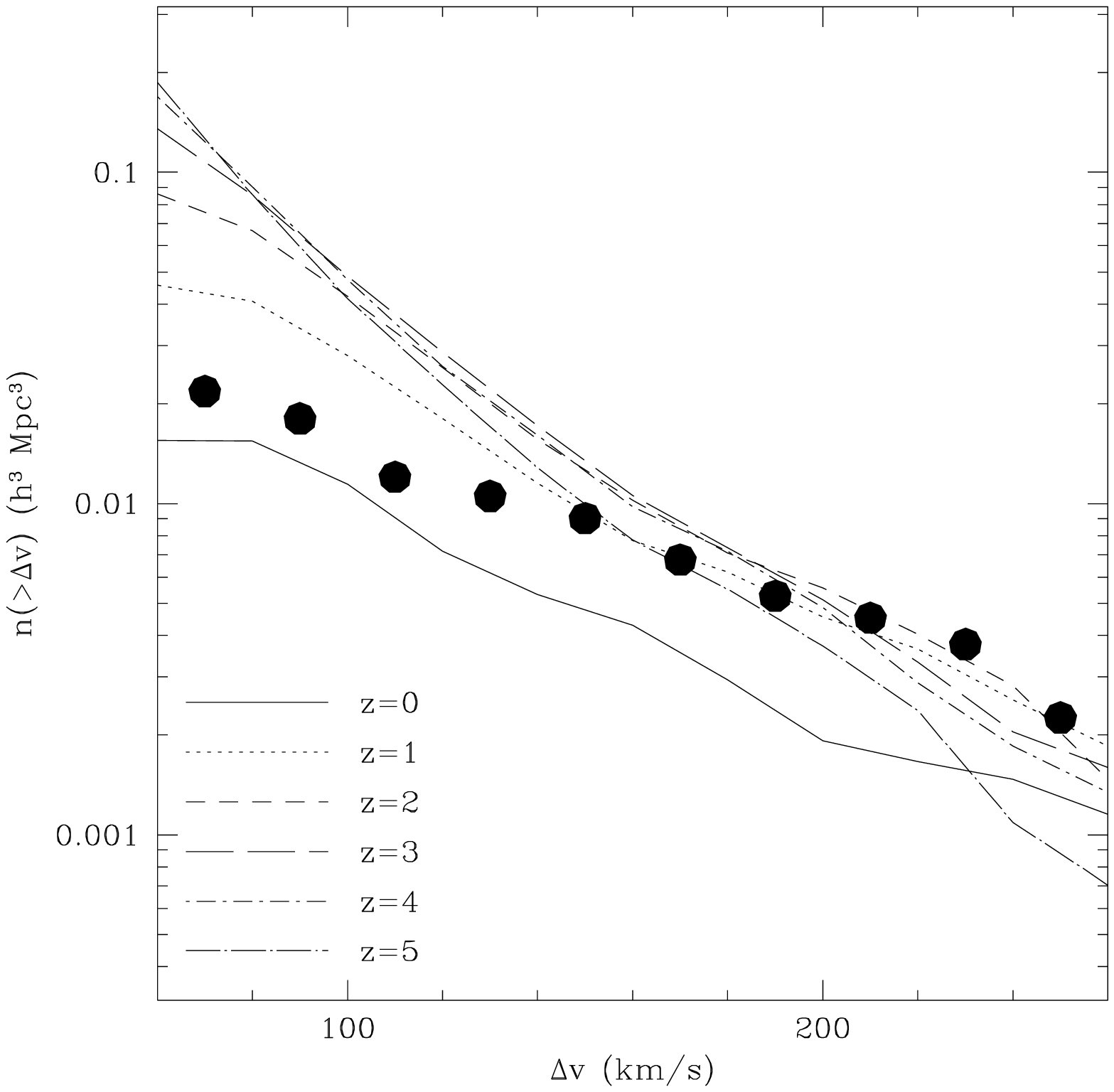}
\caption{
shows the cumulative
number density of DLAs as a function of velocity width
for six different epochs, $z=0,1,2,3,4,5$ (curves).
The solid dots indicate observations at $z>1.5$ (Prochaska \& Wolfe 2001).
}
\label{fig10}
\end{figure}
Figure 10 shows the cumulative
number density of DLAs as a function of velocity width
at $z=0,1,2,3,4,5$ (curves).
The velocity width is defined as $0.67v_{\rm rot}$, where
as $v_{\rm rot}$ is the rotation velocity of the 
DLA galaxy at radius $100h^{-1}$kpc.
The agreement between the simulation
and observations is reasonably good at the high width end
but the disagreement at the low width end is quite severe,
where a factor of $2-4$ overabundance in the simulation
is seen.
This discrepancy was seen earlier in the column density distribution 
in Figure 5, which, as we suggested,
may indicate excessive small-scale power in the adopted model.
Put aside the difference for a moment,
it is clear that DLAs are composed of galaxies with 
a broad range of velocity structures.
Our simulation says that 
most of DLAs are due to small, relatively isolated galaxies.
%(York \etal 1986; Matteucci, Molaro, \& Vladilo 1997)
In the redshift range where comparisons can be made
the results found here 
are in good agreement with extant observations
(Steidel \etal 1995;
Lanzetta \etal 1997;
Le Brun \etal 1997;
Miller, Knezek, \& Bregman 1999;
Bunker \etal 1999;
Cohen 2000;
Kulkarni \etal 2000).
%For example, at low redshift, Lanzetta \etal (1997) find a
%DLA galaxy at $z=0.16$ of luminosity
%$L_B^*/2.3$; Miller, Knezek, \& Bregman (1999)
%find a DLA galaxy at a distance of $1117$km/s of luminosity
%$L_B^*/3$;
%Cohen (2000) finds a DLA galaxy at $z=0.22$ of 
%luminosity $L_R^*/7$ and constrains another at $z=0.091$
%of luminosity $<L_R^*/30$.
%At intermediate redshift, observed DLAs appear to 
%correspond to sub-$L^*$ galaxies 
%(Steidel \etal 1995; Le Brun \etal 1997).
%Steidel \etal (1995) find a DLA galaxy
%at $z=0.86$ of luminosity $L_B^*/4$.
%At high redshift, while there is lack of directly
%identified DLA galaxies,
%indications are that they are small and lack of active star formation
%(Bunker \etal 1999; Kulkarni \etal 2000a).
However, this finding is not at odds with local 21cm observations, which
find most of the neutral hydrogen gas in the most massive galaxies.
% (Rao \& Briggs 1993).
To clarify, panel (a) of Figure 11 shows the cumulative fraction of neutral
hydrogen mass as a function of galactic stellar mass.
Panel (b) of Figure 11 shows
in a slightly different way the same information:
the solid curve indicates 
the median galaxy stellar mass 
and the two dashed curves indicate the quartile points
of galaxy stellar masses with regard to the fraction of total neutral
hydrogen mass.
Indeed, we find that most of the neutral hydrogen gas mass
at $z=0$ is in large (spiral) galaxies;
$\sim 70\%$ of neutral hydrogen gas mass at $z=0$ is 
in $L_*$ or larger galaxies, consistent with observations (Zwaan \etal 2001).

\begin{figure}
\plotone{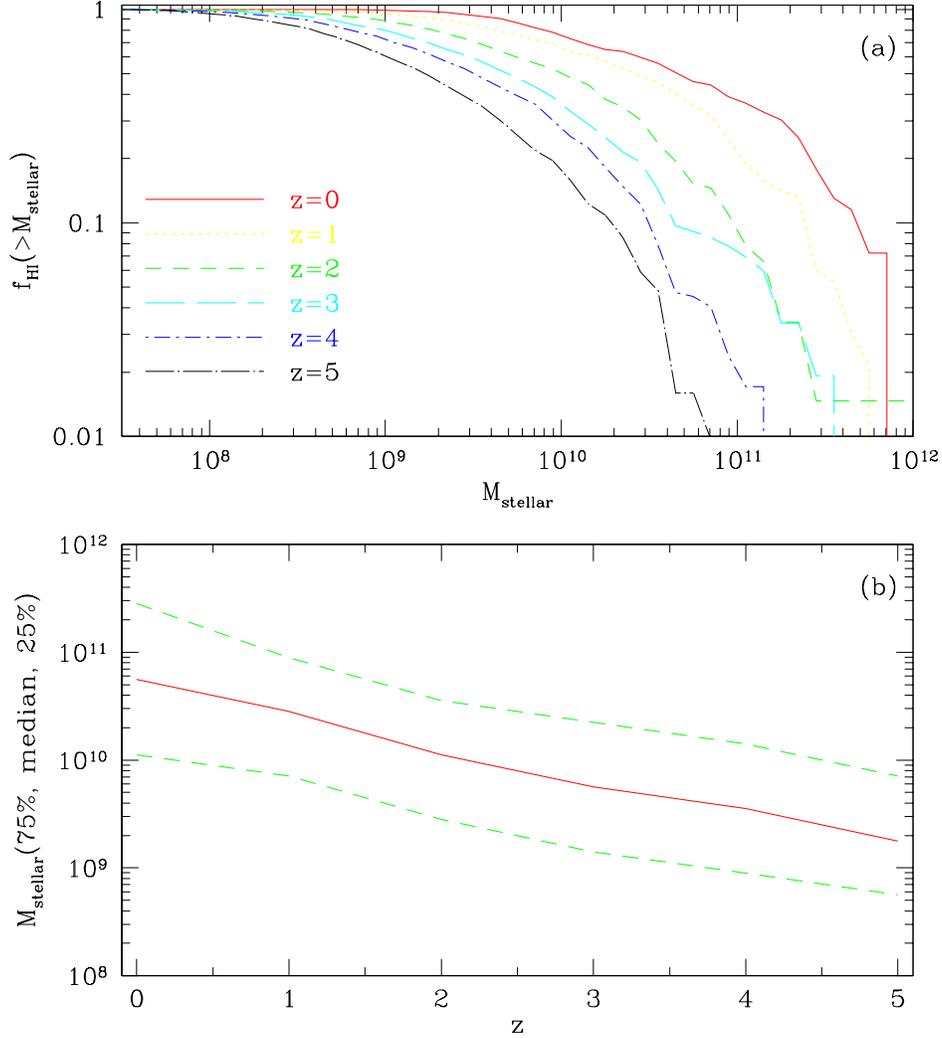}
\caption{
(a) shows the cumulative fraction of neutral hydrogen gas mass
as a function of galactic stellar mass, for six different epochs.
(b) shows the median galaxy stellar mass (solid curve)
with 50\% of the total neutral 
hydrogen mass being in galaxies more massive than that and 
the other 50\% being in galaxies less massive than that,
and the two dashed curves indicate the quartile points
of galaxy stellar masses with regard to the fraction of total neutral
hydrogen mass.
}
\label{fig11}
\end{figure}

\begin{figure}
\plotone{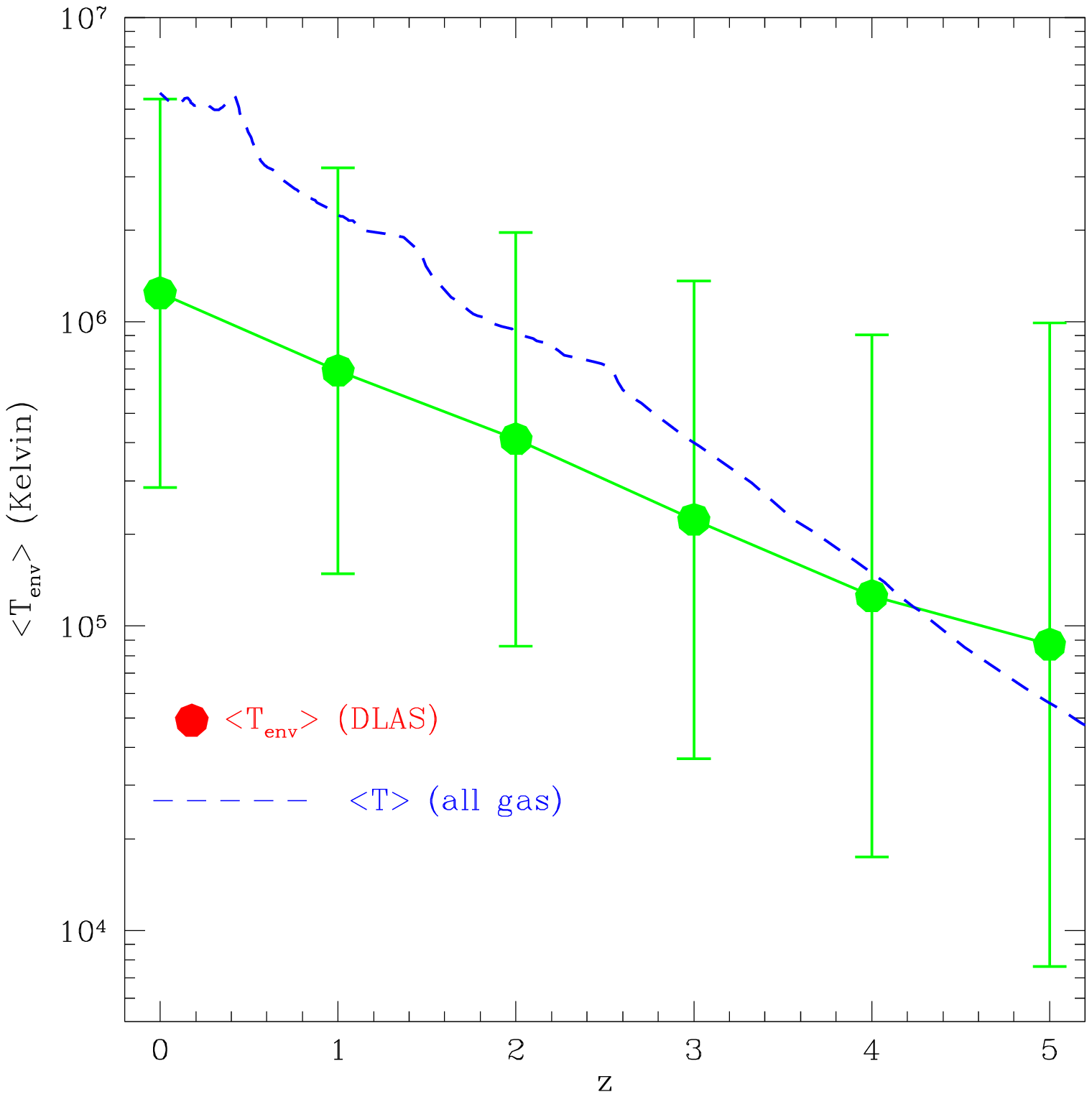}
\caption{
shows the average ``environmental temperature" of 
DLAs as a function of redshift.
%Indicated as dotted horizontal line is the actual temperature
%of the neutral gas in DLAs.
}
\label{fig12}
\end{figure}

What are the environments surrounding DLAs?
Figure 12 shows the average ``environmental temperature" in the vicinity of 
DLAs, defined as the average temperature
over 124 cells surrounding the central cell of a DLA,
as a function of redshift. 
Note that the actual (internal) temperature of the DLA gas
is low (at $<10^4~$K).
We see that
DLAs tend to reside in a higher temperature environment at lower redshift,
close to $L^*$ galaxies at $z=0$ but are in low
density regions and filaments at high redshift.
Clearly, it is unlikely that DLAs should be found in
environments like groups or clusters of galaxies at any redshift,
as indicated visually earlier in Figure 2 for $z=0$.
We show in Figure 13 the probability distribution of 
the local mass overdensity
(with a Gaussian smoothing window of radius $1h^{-1}$Mpc)
of all galaxies and DLA galaxies at three redshifts.
Consistent with Figures (2,12),
DLAs at $z=0$ 
avoid the highest density regions of clusters of galaxies
(top panel of Figure 11) as well as very low density regions,
while all galaxies extend to both
high density regions of clusters of galaxies and very low density regions.
At high redshift, DLAs still avoid very low density regions 
but not the high density regions,
because the highest density regions do not correspond 
to rich clusters of galaxies.
Figure 14 shows the proper peculiar velocity distribution 
of all galaxies and DLA galaxies at $z=0,2,4$.
In full agreement with Figure 13, DLA galaxies tend to be
found in relatively cold (low velocity) regions of filaments.

\begin{figure}
\plotone{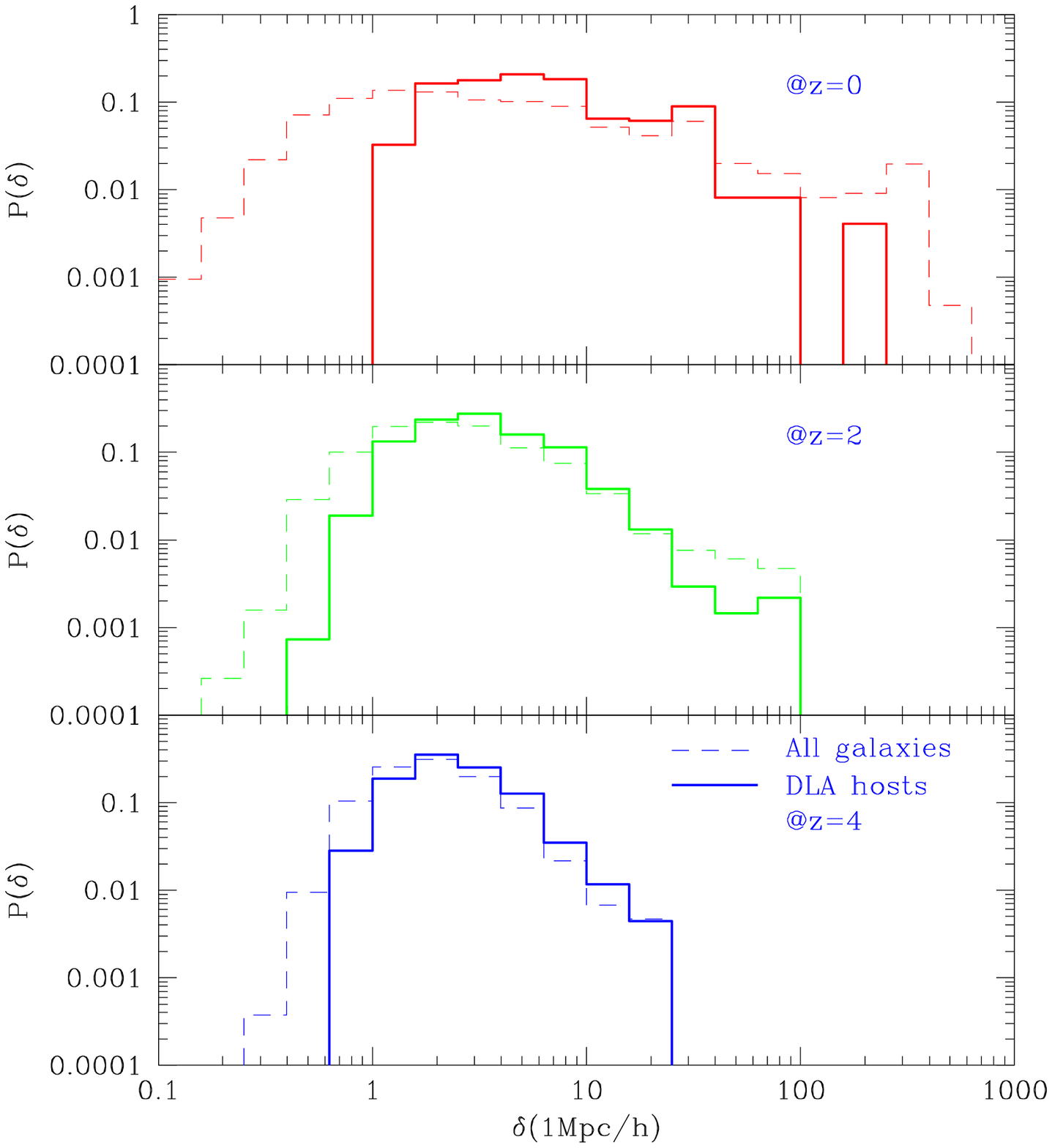}
\caption{
shows the probability distribution function of the local
overdensity of all galaxies (thin dashed histogram)
and DLA galaxies (thick solid histogram), respectively
at redshift $z=0,2,4$.
Note that at $z=0$ the DLA systems avoid both the highest
and lowest density regions.
}
\label{fig13}
\end{figure}

\begin{figure}
\plotone{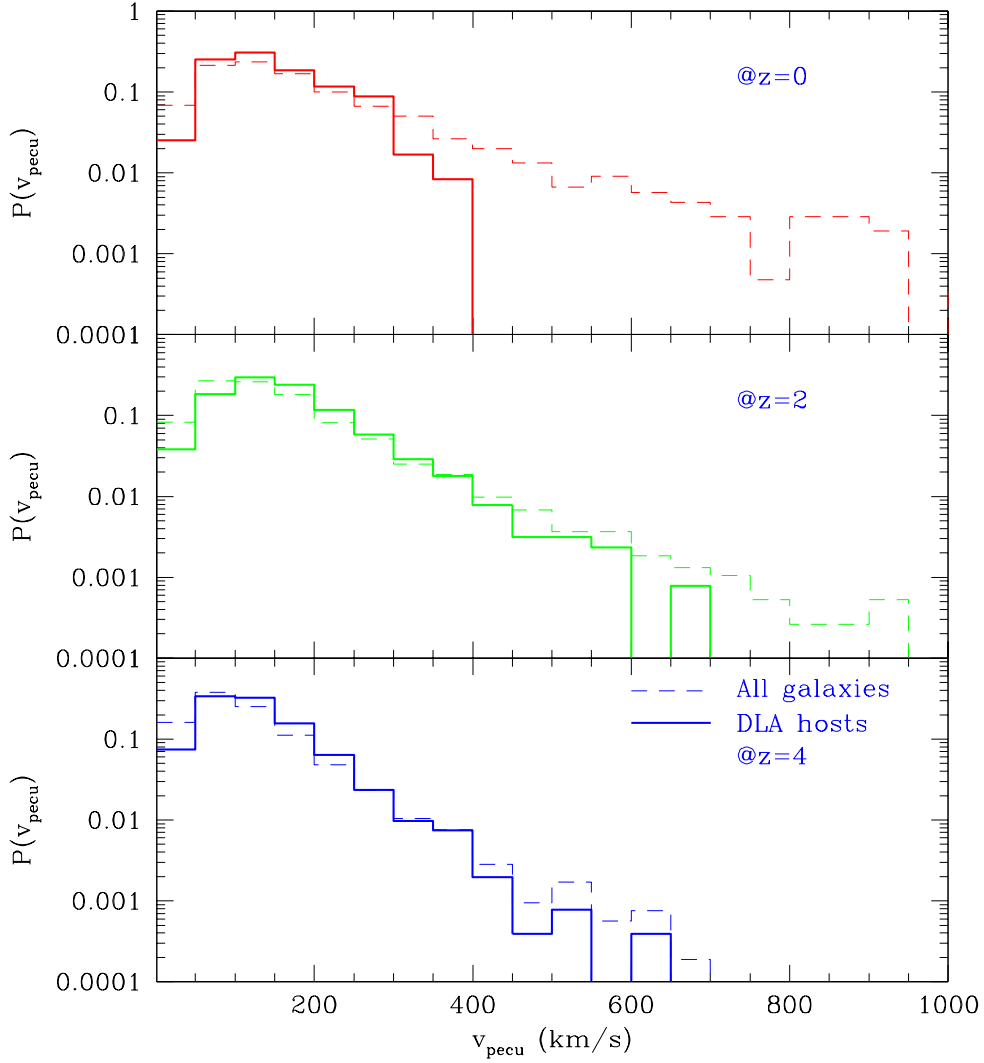}
\caption{
shows the probability distribution function of the proper peculiar 
velocity of all galaxies (thin dashed histogram)
and DLA galaxies (thick solid histogram), respectively
at redshift $z=0,2,4$.
At low redshift the DLA galaxies are in relatively
quiescent regions.
}
\label{fig14}
\end{figure}

\begin{figure}
\plotone{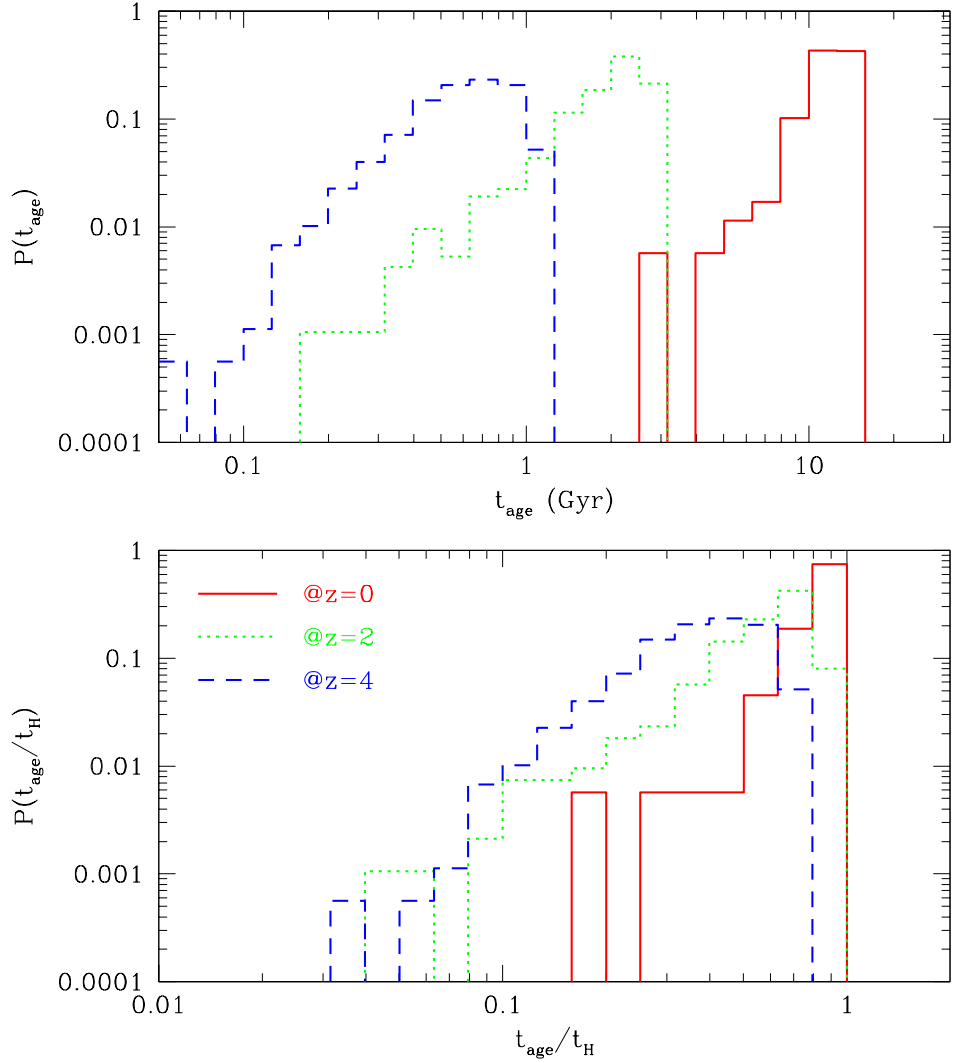}
\caption{
Top panel shows the distribution of the absolute ages of DLAs
at three redshifts (z=0,2,4).
Bottom panel shows the distribution of 
the relative age to the age of the Universe at
the respective redshifts.
}
\label{fig15}
\end{figure}

\begin{figure}
\plotone{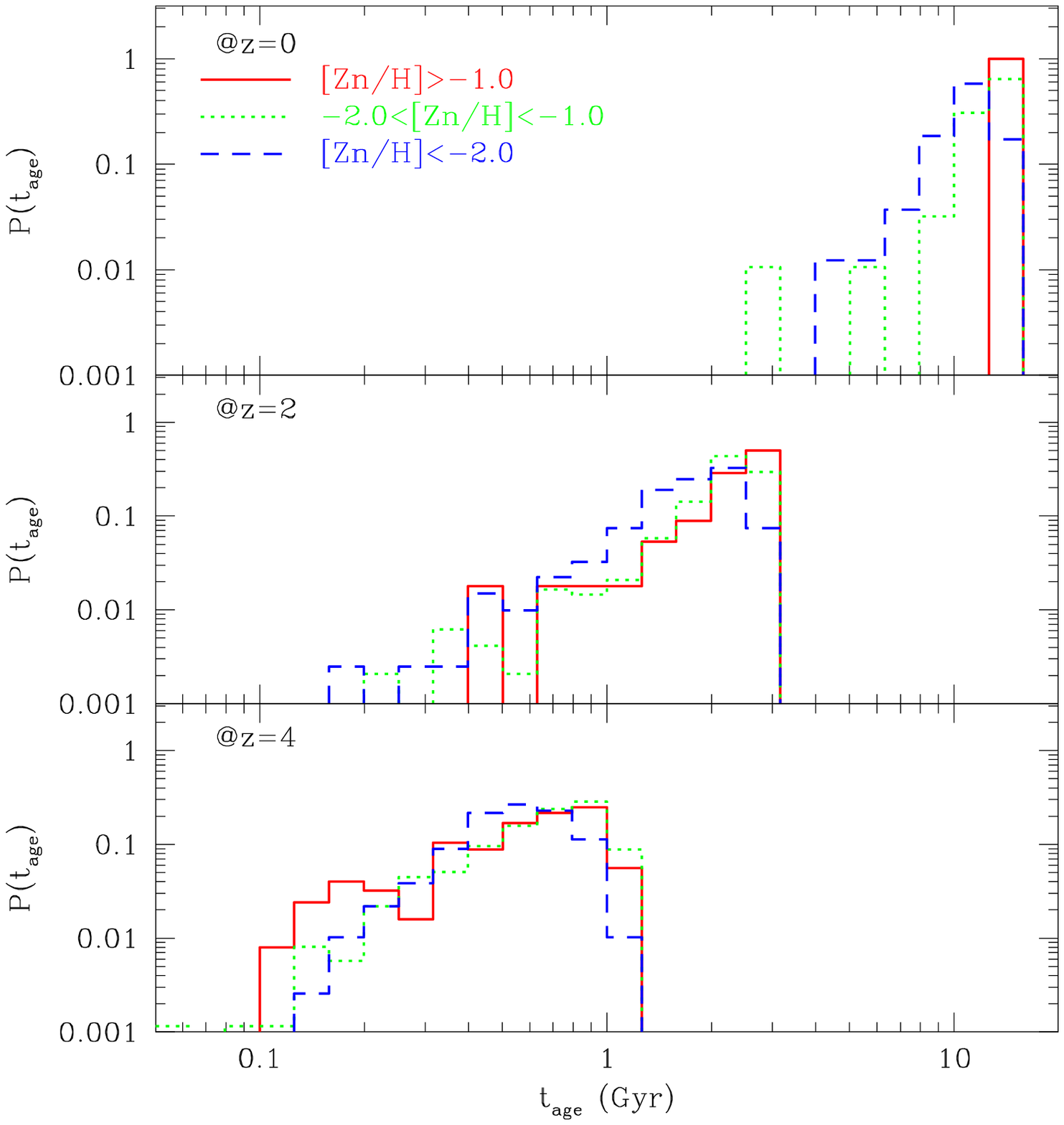}
\caption{
shows the age distributions for three subsets
at three different redshifts.
We divide the metallicities of DLAs into 
three subsets: $[Zn/H]>-1.0$, $-2.0<[Zn/H]<-1.0$ and $[Zn/H]<-2.0$.
}
\label{fig16}
\end{figure}

How old are DLA host galaxies?
Figure 15 shows the absolute ages of DLAs (top panel)
and the relative age (to the age of the Universe) (bottom panel)
at three redshifts (z=0,2,4).
We see that DLAs at $z=0$
are old with their ages close to the age of the universe
and no DLAs are expected to be younger than $5~$Gyr.
Going to high redshifts DLAs become relatively younger
and at $z=4$ the median age of DLAs 
is about half the age of the universe at that redshift.
In addition, the spread in age distribution becomes
larger at high redshifts with the full width
spanning about one decade at $z=2-4$. 
We then subdivide the DLAs according to their metallicity
and compute separately the age distributions of the subsets.
Figure 16 shows the age distributions for three subsets
at three different redshifts.
We see that at $z=0$ a trend is clearly visible:
high metallicity DLAs are older
than their low high metallicity counterparts
and no DLAs with $[Zn/H]>-1.0$ are expected to be younger
than $10$Gyr at $z=0$.
However, at high redshift ($z=2-4$) there does not appear to 
exist any significant age segregation with respect to DLA
metallicity.

The fact that a typical DLA galaxy, nearly independent of its
metallicity, would be $\sim 0.8-2$Gyr old at $z=2-4$ 
is interesting in light of recent observations.
Pettini \etal (2002) find that, based on 
a high resolution, high S/N sample of 15 DLAs,
the nitrogen abundance in DLAs with $[O/H]=-2$ to $-1$
has not saturated by $<z>=2.4$.
The indicated age of DLAs at the relevant redshift in our model, 
if taken to interpret the observational data,
would imply a requirement that the release of nitrogen
from intermediate mass stars be delayed by $\sim 1$Gyr,
the typical age of DLAs derived here.
If the delay is much shorter than 
the typical age of DLAs,
then the nitrogen abundance in DLAs would have already saturated.

The ages of DLA galaxies appear to be 
comparable to those of LBGs (see Figure 10a of Shapley et al. 2001).
It may suggest that LBGs have higher star formation rate 
than DLA galaxies, which, of course, is entirely consistent
with the lower metallicity of DLAs.

Finally, Figure (17) 
shows the fractions of metal mass in gaseous phase
as a function of redshift,
divided into three components:
DLAs (with temperature $T< 10^4$K; solid curve),
intracluster gas ($T>10^7$K; dashed curve),
and the remaining IGM gas ($10^4\le T\le 10^7$K; dotted curve).
We see that at $z\ge 1$
roughly one half of the metals in the gas phase
is in DLAs and the other half in the general 
intergalactic medium, while the fraction of
the metals in very hot, rich clusters is small.
But by $z=0$ only about 25\% of metals are in DLAs,
while the fraction in clusters has increased dramatically,
reaching $\sim 30\%$, consistent with the fact
that most clusters form at $z<1$ in this model.
The decline of metal mass fraction in DLAs at low redshift
is partly due to metal-enriched gas being 
collected in the intra-cluster gas and partly
due to star formation out of metal rich gas 
at low redshift (as indicated by the long-dashed curve).
Our calculations indicate that
roughly $40\%$ of all metals are initially ejected
into the IGM through supernovae and then 
re-formed into and locked up in stars by $z=0$;
this fraction is lower at $20\%$ at $z=3-5$.

\begin{figure}
\plotone{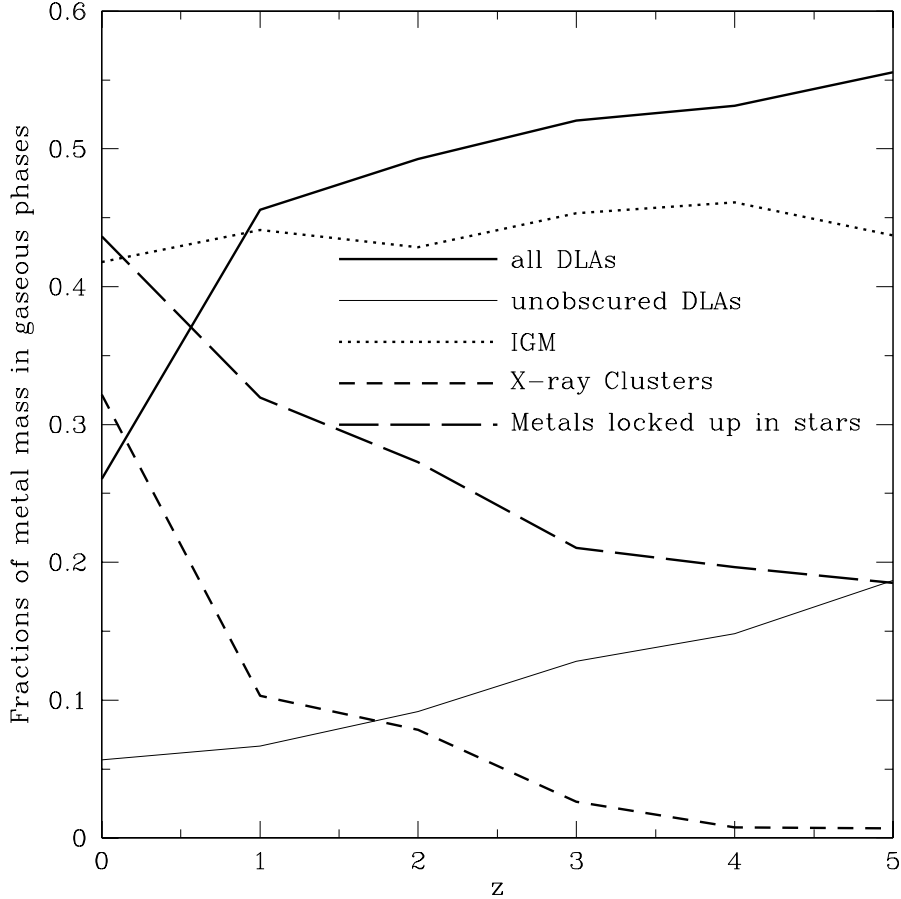}
\caption{
shows the fractions of metal mass in the {\it gaseous phase}
(i.e., not including metals in stars)
divided into three components at each epoch:
DLAs (with temperature $T< 10^4$K; solid curve),
intracluster gas ($T>10^7$K; dashed curve),
and the remaining IGM gas ($10^4\le T\le 10^7$K; dotted curve).
Also shown as long-dashed curve is the fraction of metal mass
in stars (i.e.,
metals previously ejected into IGM and then recycled back 
into stars) in terms of the {\it total} metal mass 
(that has been synthesized through stars and ejected back into IGM 
at one point in time at least).
}
\label{fig15}
\end{figure}
We note that most of the metals in DLAs are in the dust ``obscured"
high metallicity, high column density DLAs, according
to our model (see Figures 5,6 and related discussion).
Nevertheless, the fraction of total 
metals in the ``unobscured" DLAs (thin solid curve in Figure 13)
ranging from $\sim 5\%$ at $z=0$ to
$\sim 20\%$ at $z=5$, in good agreement with 
observational estimates (Pettini 1999; Pagel 2001).
If we identify the dust ``obscured"
DLA population with SCUBA galaxies at high redshift,
the implication is that SCUBA galaxies may contain 
$\sim 70\%$ of metals residing in galaxies 
or $\sim 40\%$ of all metals at high redshift,
which appears to be in agreement with recent observations
(Dunne \etal 2002).
In addition, since both 
the computed metallicity ($\sim \zsun/3$) and
abundance of X-ray clusters at low redshift are found to 
be in agreement with observations,
the computed fraction of metals in the intracluster gas
should be approximately correct.

Before concluding the paper, one word on numerical resolution 
is appropriate.
Since the real DLAs have sizes comparable to our numerical resolution
at $z=0$ (30kpc comoving; note that our numerical resolution 
is better at high redshift scaling as 30kpc/(1+z) kpc proper),
it seems inevitable that our limited numerical resolution
must have affected the results.
In particular, the actual density and hence the column density
of the computed DLAs may have been underestimated.
The countervailing effect is that lower resolution 
may artificially enlarge the central dense regions
thus produce more DLAs (if the central density is sufficiently high).
Therefore, the net effect is not immediately obvious.
In any case, to test the resolution 
effect we artificially lower the
column density  threshold
that we use to identify DLAs 
from $2\times 10^{20}$cm$^{-2}$ 
to $1\times 10^{20}$cm$^{-2}$.
We find that this change does not change 
the results presented in all the figures noticeably,
except that in Figure 5 the unweighted metallicity (thin solid curve)
of all DLAs is slightly lower by $0.1$ dex at $z=3$ 
and $0.2$ dex at $z=4$.

We should still stress that we do have enough
resolution to resolve the internal structures of galaxies.
All that we are trying to present 
in the paper are global properties averaged over 
all galaxies and intergalactic
medium on scales comparable to or 
larger than our resolution, that do not sensitively 
depend on details on the sub-grid scales. 
To support our view that our treatment of star formation and metal production 
rates are reasonable and empirically in agreement with observations over a 
wide range of observables, we will refer to published papers based on the 
same simulation (Cen \& Ostriker 1999; Nagamine \etal 2000;
Nagamine \etal 2001a,b).
To summarize, there are at least three lines of evidence that would
indicate that the simulation is adequate for tracking global star
formation, galaxy formation and metal production.
First, we have computed the global star formation rate as a function
of redshift (Nagamine etal 2000, 2001a;), i.e., the ``Madau plot", and found
reasonable agreement with observations over the redshift range that can be
compared, given the uncertainties concerning both observations and simulations.
This indicates that the observed star formation history is reasonably
well produced by the simulation of this currently best cosmological model.
This would remove most of the uncertainties in the simulation concerning 
the amount of stars formed. Second, the computed luminosity function of
galaxies are found to be in reasonable agreement with observations with a 
flat faint end slope of $-1.15$ down to a galaxy mass of 
$10^{8.5}\msun$; Nagamine etal 2001b). 
Once again, this indicates that our simulation has
sufficient resolution as to identify galaxies and compute the mass and 
luminosity of galaxies (but not to resolve internal structures of galaxies).
Third, the resulting metallicity of the computed intra-cluster gas 
($\sim 1/3$ solar value) is in excellent with observations, implying that
our adopted metal yield value ($0.02$, based on stellar interior
theory, see Arnett 2000) agrees with empirically determined observational
value. Since the star formation history and galaxy luminosity function
are independent of metal yield and most of the metals ending up in 
the intracluster medium are due to galaxies close to those hosting
DLAs, the yield is essentially fixed, on which metallicity of DLAs mainly 
depend. In addition, our previous, lower resolution simulations 
(Cen \& Ostriker 1999) 
have yielded results over a wide range of environments (from Lyman
alpha forest to clusters of galaxies) that are in reasonable agreement
with observations, further supporting the conclusion that our overall 
metal production rates as well as those regions previously resolved
has been correctly computed.

\section{Conclusions}

We use a latest high mass resolution 
hydrodynamic simulation of a $\Lambda$CDM model
to compute the metallicity evolution of damped Lyman alpha systems.
Contrary to a naive expectation,
but consistent with observations,
the computed true mean metallicity of all  
DLAs shows a very slow evolution with time.
Observations appear to pick out damped
Lyman alpha systems to be at an intermediate stage
in galactic evolution.
They tend to live in the filaments,
where metal enriched gas from previous
generations of small star forming subgalactic units has
collected via the hierarchical merger process.
As star formation and merging proceeds,
they tend to turn into normal massive galaxies ultimately 
often residing in high
density high temperature regions where the gas
component is stripped off.
%Thus the internal metallicity of systems identified as DLAs
%is typically higher than that averaged over all (mostly small) 
%galaxies, lower than that in massive galaxies
%and evolves slower than followed by the typical
%Lagrangian gas mass element.

It is important to note that 
DLAs are not a simple population of galaxies 
but due to sightlines through 
a variety of different systems.
The mixture of galaxy hosts of DLAs changes with redshift.
The median luminosity of a DLA, $L_{DLA}(z)$,
in units of typical galaxy luminosity  at that redshift,
$L^*(z)$, that is, $(L_{DLA}/L^*)_z$ decreases
from $1.1$ to $0.5$ as redshift declines from $z=3$ to $z=0$,
but the absolute luminosity of the median DLA system
increases in the same interval by a
factor of $5$ from $0.1L^*(z=0)$ to $0.7L^*(z=0)$.
We predict that DLAs should not be found in hot environments
such as groups and clusters of galaxies,
but tend to live in regions moderate overdensity with relatively
cold flows.

The age of the DLA host galaxies approaches the age of the universe at $z=0$,
while it is approximately half the age of the universe
at $z=3-4$.
At high redshift ($z=2-4$) there appears to 
exist no significant age segregation with respect to DLA metallicity.

We find that about 50\% all metals in gaseous phase
is in DLAs from $z=5$ to $z=1$,
then decreasing rapidly to $\sim 25\%$ by $z=0$,
as these metals are swept into hotter
X-ray emitting cluster gas and recycled into newly formed stars.
It is noted that about 50\% of total metals are locked
up in stars by $z=0$; this number is 20\% at $z\ge 3$.

Observational selection effects
(such as dust obscuration)
may have significantly masked the true average metallicity of DLAs.
Using only those DLAs that
are located in the unobscured region of DLAs 
in the metallicity-column density plane (using a very simple
dust obscuration model),
we find that, while the trend of metallicity evolution remains little
unchanged, the mean (column density-weighted) metallicity
is reduced by 0.3 to 0.5 dex from $z=0$ to $z=5$,
bringing the simulation results
into good agreement with observations.
The ``unobscured" DLAs in the simulation also provide
good matches to both the observed column density distribution,
redshift evolution of the neutral gas content in DLAs and
metal content in DLAs.
Therefore, in order for the model to 
agree with observations for a range of variables
either the assumed dust obscuration
effect occurs in the real universe or the model has to be
modified as to remove the high column density, high metallicity DLAs. 
The possibilities for the latter will be explored in the future.
It should be emphasized, however, 
that current observations do not seem to 
require significant amount of dust in DLAs 
(Ellison \etal 2001; Prochaska \& Wolfe 2002).
We point out, however, that only $1.8$ DLAs
are expected to be dust obscured (in optical)
in the CORALS radio survey of Ellison \etal (2001) of total $19$ DLAs.
Clearly, a sample several times larger is required 
to greatly firm up the statistical significance of 
their important findings and determines the importance of dust
obscuration unambiguously.
While our simulations strongly suggest 
the significance of dust obscuration with regard to DLA evolution,
we will be forced to 
make a substantial revision of the computed model,
if significant dust obscuration is firmly excluded 
observationally in the future.
We stress, however, that some of the major conclusions
of the present paper, including
the mild metallicity evolution of DLAs, the large
disparity of DLA host galaxies and the monotonic evolution 
of the increasing mass of DLA systems with time, 
are much less prone to these uncertainties and
not expected to alter significantly.

Finally, we pointed out that 
that the adopted model predicts
(approximately a factor of $2-3$)
too many damped Lyman alpha systems,
indicating either imperfections of the simulation,
or the possibility that this 
$\Lambda$CDM model has too much small scale power.

\acknowledgments
The work is supported in part
by grants AST93-18185 and ASC97-40300.
We thank K. Nagamine for kindly
providing the catalog of simulated galaxies,
Greg Bryan for kindly making a software program available to us
which is used to produce Figure 1,
and Max Pettini and Mike Fall for very useful comments.

\newpage

\end{document}